\documentclass[]{aa}

\usepackage{graphicx}
\usepackage{txfonts}

\usepackage{hyperref}
\hypersetup{
    colorlinks,
    linkcolor={blue},
    citecolor={blue},
    urlcolor={blue}
}
\usepackage[T1]{fontenc}

\usepackage{amsmath}    
\usepackage{amssymb}    
\usepackage{xcolor}     
\usepackage{booktabs}       
\usepackage{here}

\usepackage{float}
\usepackage{placeins}
\usepackage{lipsum}
\usepackage{tabularx}
\usepackage{newtxtext,newtxmath}
\usepackage{multirow}

\usepackage[T1]{fontenc}

\usepackage{multicol}
\usepackage{graphicx}
\usepackage{multirow}

\usepackage{subcaption}

\usepackage{color}

\DeclareRobustCommand{\VAN}[3]{#2}
\let\VANthebibliography\thebibliography
\def\thebibliography{\DeclareRobustCommand{\VAN}[3]{##3}\VANthebibliography}

\usepackage{graphicx}   
\usepackage{amsmath}    

\usepackage{threeparttable}

\newcommand{\bbeta}{$B_{\beta=50}$}

\usepackage{soul}
\graphicspath{{Figures/}}
\newlength{\fullwidth}
\newlength{\halfwidth}
\begin{document}

   \title{Simulating the LOcal Web (SLOW) \\II. Properties of local galaxy clusters}

   \author{Elena Hernández-Martínez\thanks{\email{hernandez@usm.lmu.de}}\fnmsep
          \inst{1}
          \and
          Klaus Dolag\inst{1,2}
          \and
          Benjamin Seidel\inst{1}
          \and 
          Jenny G. Sorce\inst{3,4,5}
          \and\\
          Nabila Aghanim\inst{4}
          \and
          Sergey Pilipenko\inst{9}
          \and
          Stefan Gottl{\"o}ber\inst{5}
          \and
          Th\'eo Lebeau\inst{4}
          \and
          Milena Valentini\inst{1,6,7,8}
          }

   \institute{Universit\"ats-Sternwarte, Fakult\"at f\"ur Physik, Ludwig-Maximilians-Universit\"at M\"unchen, Scheinerstr.1, 81679 M\"unchen, Germany
         \and
             Max Planck Institute for Astrophysics, Karl-Schwarzschild-Str. 1, D-85741 Garching, Germany
\and
Univ. Lille, CNRS, Centrale Lille, UMR 9189 CRIStAL, F-59000 Lille, France
\and
Universit\'e Paris-Saclay, CNRS, Institut d'Astrophysique Spatiale, 91405, Orsay, France
\and
Leibniz-Institut f\"{u}r Astrophysik (AIP), An der Sternwarte 16, D-14482 Potsdam, Germany
\and
Astronomy Unit, Department of Physics, University of Trieste, via Tiepolo 11, I-34131 Trieste, Italy
\and
ICSC - Italian Research Center on High Performance Computing, Big Data and Quantum Computing
\and
INAF – Osservatorio Astronomico di Trieste, via Tiepolo 11, I-34131 Trieste, Italy
\and
P.N. Lebedev Physical Institute of the Russian Academy of Sciences, Profsojuznaja 84/32, Moscow 117997, Russia
             }

   \date{Accepted XXX. Received YYY; in original form ZZZ}

 
  \abstract
   {This is the second paper in a series presenting the results from a 500 $h^{-1}$Mpc large constrained simulation of the local Universe (SLOW). The initial conditions for this cosmological hydro-dynamical simulation are based on peculiar velocities derived from the CosmicFlows-2 catalog. The simulation follows cooling, star formation, and the evolution of super-massive black holes. This allows one to directly predict observable properties of the intracluster medium (ICM) within galaxy clusters, including X-ray luminosity, temperatures, and the Compton-y signal.}
   {Comparing the properties of observed galaxy clusters within the local Universe with the properties of their simulated counterparts enables us to assess the effectiveness of the initial condition constraints in accurately replicating the mildly nonlinear properties of the largest, collapsed objects within the simulation. }
   {Based on the combination of several, publicly available surveys we compiled a sample of galaxy clusters within the local Universe, of which we were able to cross-identify 46 of them with an associated counterpart within the SLOW simulation. We then derived the probability of the cross identification based on mass, X-ray luminosity, temperature, and Compton-y by comparing it to a random selection. }
   {Our set of 46 cross-identified local Universe clusters contains the 13 most massive clusters from the \textit{Planck} SZ catalog as well as 70\% of clusters with $M_{500}$ larger than $2\times 10^{14}$ M$_{\odot}$. Compared to previous constrained simulations of the local volume, we found in SLOW a much larger amount of replicated galaxy clusters, where their simulation-based mass prediction falls within the uncertainties of the observational mass estimates. Comparing the median observed and simulated masses of our cross-identified sample allows us to independently deduce a hydrostatic mass bias of $(1-b)\approx0.87$.}
   {The SLOW constrained simulation of the local Universe faithfully reproduces numerous fundamental characteristics of a sizable number of galaxy clusters within our local neighborhood, opening a new avenue for studying the formation and evolution of a large set of individual galaxy clusters as well as testing our understanding of physical processes governing the ICM. }


   \maketitle
%




\section{Introduction}
\label{Introduction}

Extremely interesting times for cosmology and astrophysics are ahead of us. 
Within the past years, tremendous efforts have been put into developing missions to 
achieve larger cosmological surveys, for example the Sloan Digital Sky Survey (SDSS; \citealt{2019BAAS...51g.274K}), the Dark Energy Survey (DES; \citealt{2016MNRAS.460.1270D} ), and the 2df Galaxy Redshift Survey (2dFGRS; \citealt{2001MNRAS.328.1039C}{}{}), as well as deeper local surveys, for example the Arecibo Legacy Fast ALFA Survey(ALFALFA; \citealt[][]{2007AJ....133.2087S}{}{}), the Two Micron All Sky Survey (2MASS; \citealt{2006AJ....131.1163S}{}{}), and the Nearby Galaxies Legacy Survey (NGLS; \citealt[][]{2009ApJ...693.1736W}{}{})).

The progress in this field has been accompanied by the constant development of 
higher resolution hydrodynamical cosmological simulations \citep[e.g.,][]{2004ApJ...615L.101B, 2014Natur.509..177V,Hirschmann_2014,Dolag_2016, Dubois_2016}, which have become a fundamental theoretical tool in the study of the multi-scale and nonlinear nature of astrophysical and cosmological processes \citep[][]{Overzier_2016}{}{}.

Among the structures under study in our Universe, galaxy clusters represent a 
fundamental set of cosmological probes. Lying at the intersection 
between astrophysics and cosmology, the understanding of the formation and evolution
of these structures is primordial in our comprehension of the Universe, spanning from the scale of particle physics, such as in the study of dark matter candidates  \citep[e.g.,][]{2000NJPh....2...11T, 2009PhRvD..79k5016K, 2019PhRvD.100c5001M}{}{}, to the largest cosmological scales when, for example, aiming to tight the constraints on cosmological parameters \citep[e.g.,][]{ 2010MNRAS.406.1759M,  2015MNRAS.446.2205M, 2016ApJ...832...95D,  2017MNRAS.471.1370S, 2018A&A...620A..10P, 2020MNRAS.499.4768I, 2023MNRAS.522.1601C}{}{}

Nevertheless, retrieving detailed information about galaxy clusters through observations 
is indirect and may suffer from biases.
Thus, the use of cosmological simulations has turned out to be a unique complementary tool 
in the study of the formation of these structures and to assess possible biases in the observations \citep[see e.g.,][for more extensive reviews]{2012ARA&A..50..353K, frenk12,2020NatRP...2...42V}{}{}.

Traditionally, what we call standard cosmological simulations of structure formation, such as Magneticum \citep{Hirschmann_2014, Dolag_2016}, Horizon-AGN \citep[][]{Dubois_2016}{}{}, EAGLE \citep{2015MNRAS.446..521S}, and IllustrisTNG \citep[][]{2018MNRAS.473.4077P}{}{}, have aimed to produce "random" statistically representative patches of the Universe. Even if these simulations can reflect properties of our Universe in a statistical manner, such as large-scale structure, clustering statistics, and galaxy diversity \citep[e.g.,][among others]{2011ApJ...740..102K, 2012MNRAS.423.3018P, 2012MNRAS.426.2046A, Watson_2014, 2014ApJS..210...14K,skillman2014dark}, they do not contain the specific objects (such as the Virgo and Coma clusters, or the local voids), embedded within the correct large-scale structure. Thus, there remains an underlying difference between the ensemble mean prediction and the observational dataset, which is subject to cosmic variance  \citep[e.g.,][]{2002AJ....123..485S, 2003AJ....126.2081A, 2009ApJS..182..543A} .  

To determine the nature of specific objects in our Universe, many studies have used large random simulations in order to search for analogs that are similar to the particular object in question (e.g., ELVIS \citealt{2014MNRAS.438.2578G, 2019MNRAS.487.1380G} and APOSTLE \citealt{2016MNRAS.457.1931S} projects). This approach is useful for assessing the frequency of certain types of objects within the $\Lambda$CDM framework, but it falls short in accurately defining the nature of these objects. This is because the nonlinear process of structure formation in the Universe allows many evolutionary routes to reach a seemingly similar end state. To truly understand the nature of any specific observed object, it needs to be modeled within its accurate cosmological context, not just a random one. This is the exact aim of "constrained simulations" of the Universe.

Constrained simulations are created using a set of $\Lambda$CDM initial conditions constrained with actual galaxy observations, designed to develop into the specific distribution of large-scale structures seen in the local observed Universe \citep[e.g.,][for a nonextensive list]{2002ApJ...571..563K,2002MNRAS.333..739M,2004JETPL..79..583D, 2014MNRAS.437.3586S, sorce2016cosmicflows, 2016MNRAS.458..900C,  2020MNRAS.496.5139S, 2022MNRAS.515.2970S, 2022MNRAS.512.5823M, Dolag_2023, 2023MNRAS.523.5985P}. This approach ensures that the entire array of structures in the local area, including galaxy clusters, filaments, and voids, is accurately replicated in their respective locations. Through this method, we can replicate not only the positions of these local structures but also their velocities, masses, and internal characteristics. Additionally, it allows us to understand their distinctive formation histories and the environments in which they develop.

This method is highly beneficial because it enables us to leverage observations from our local environment, which are of comparatively unparalleled data quality. It allows us to conduct accurate direct comparisons, avoiding potential biases inherited in statistical samples, such as cool-core bias in X-ray-selected galaxy cluster samples \citep[][]{2011A&A...526A..79E, 2008A&A...491...71P, 2017ApJ...841....5N}{}{}, biases for merging clusters in weak-lensing-selected clusters \citep[][]{2020ApJ...891..139C, 2022MNRAS.515.4471W}, or the hydrostatic mass bias \citep[][]{Gruen_2014, von_der_Linden_2014, 2015MNRAS.449..685H, Applegate_2016, Okabe_2016, Sereno_2017, Hurier_2018, Medezinski_2018, Jimeno_2018}.

Even more, these simulations can serve as a valuable tool to infer potential biases inherent in statistical comparisons between extensive surveys and standard cosmological simulations (e.g., the halo bias in the local Universe, \citealt{Dolag_2023}, or the hydrostatic bias expected for the Virgo cluster, \citealt{ 2023arXiv231002326L}). They can even go beyond the statistical approach and trace back evolutionary paths of the studied structures \citep[][]{Sorce_2016, 2020MNRAS.496.5139S}. These kinds of simulations have indeed already been proven to be essential in enhancing our understanding of the properties and evolution of individual clusters within the cosmic web environment \citep[e.g.,][for a non-extensive list]{2010MNRAS.401...47D,2013A&A...554A.140P,2016A&A...596A.101P,Olchanski_2018, 2019A&A...625A..64J, 2020MNRAS.496.5139S, Sorce_2021, Malavasi_2023,2023arXiv231002326L}. 

In order to be able to represent the local Universe properly, one needs a large enough volume of several hundred megaparsec. Additionally, the initial conditions must be constrained by parameters that are not directly derived from the distribution of observable tracers. This condition ensures that the comparison remains independent. Moreover, it is essential to incorporate the presence of baryons and to model the
physics of galaxy formation within
the simulations. Once the evolution of the intracluster medium (ICM) is %
accurately reproduced, it allows one to %
identify galaxy clusters
based on observables, such as X-ray luminosities, temperatures, or the Sunyaev-Zeldovich (SZ) effect \citep[][]{1970Ap&SS...7....3S}{}{} of galaxy clusters.

Constrained simulations are inherently rooted in observations, and the quality of these simulations is contingent upon both the quality and the diversity of environment of the constraints \citep[][]{Sorce_2017}. Initially, constrained simulations relied on data obtained from redshift surveys, which estimated the local matter density by analyzing the distribution of galaxies based on their observed redshifts. An overview of projects carried out using these simulations can be found in Sect. 2 of \citet{Dolag_2023}.

A more recent method for constraining initial conditions (ICs) is grounded in peculiar velocities obtained from direct distance measurements. Tracing the peculiar velocity field involves mapping the potential field generated by the large-scale density distribution of all matter, a factor independent of the tracer population. Managing the dataset of galaxies with measured distance indicators, as gathered by the CosmicFlow project \citep[][]{2012ApJ...744...43C, 2013AJ....146...86T, 2016AJ....152...50T, Tully_2023}, requires careful handling. Techniques have been developed to accommodate for the growing number of constraints and address inherent biases in velocity data \citep[][]{Doumler_2013a, Doumler_2013b, Doumler_2013c, 2014MNRAS.437.3586S, sorce2016cosmicflows, Sorce_2015}.  
The ICs derived from this approach have been used for various simulations, including those related to cosmic reionization  \citep[][]{Ocvirk_2020}, high-resolution local group simulations \citep[HESTIA,][]{Libeskind_2020}, and dark matter simulations that successfully replicate,  for example, 
the Virgo and Coma clusters -- Constrained LOcal and Nesting
Environment Simulations \citep[CLONES;][]{Sorce_2018}. More recently, the re-simulation of the regions around these prominent clusters with the zoom-in technique has allowed for studies of galaxy properties within the Virgo and Coma
clusters, with re-simulated runs including galaxy formation physics \citep[][]{Sorce_2021}. 

The Simulating the Local Web \citep[SLOW; ][]{Dolag_2023}{}{} simulation is a 500 $h^{-1}$Mpc box, using one realization of CLONES \citep[][]{Sorce_2018}.
It follows the evolution of dark and baryonic matter representing our cosmic neighborhood and centered on the position of the Milky Way (MW). The initial conditions' constraints for these simulations are based on the direct distance measurements and redshift estimates of
several thousand galaxies from the second catalog of the Cosmicflows project \citep[ CF2,][]{2013AJ....146...86T}{}{}, see Sect. \ref{The SLOW simulations} for a more detailed description.

Moreover, baryonic matter is treated using hydrodynamics in conjunction with various state-of-the-art sub-grid models. The simulation set includes boxes incorporating galaxy physics, such as the formation of stellar populations, black holes (BHs), and associated active galactic nuclei (AGN) physics, for which the prescriptions of the Magneticum simulations\footnote{www.magneticum.org} \citep[see][]{Hirschmann_2014, Dolag_2016} were used. Additionally, we also performed a run incorporating the evolution of magnetic fields and cosmic rays (see 
Sect. \ref{The SLOW Simulation Set}). 

This second paper in the series presents 46 galaxy cluster replicas reproduced in our constrained local Universe simulation box. It provides a detailed description of the identification process of the clusters, which made use of observational multiwavelength data for a one-to-one identification. Additionally, we present the general properties of our galaxy cluster replica and evaluate the statistical significance of their match as well as their predicted observational signatures in comparison with their observed counterparts.

The structure of the paper is as follows: In Sect. \ref{The SLOW simulations}, we provide a detailed description of how the constrained initial conditions were developed, along with insights into the physics incorporated into the simulation. Moving to Sect. \ref{Local galaxy clusters}, a broad perspective on the most prominent structures within the local Universe is presented. Section \ref{Scaling Relations for Low Redshift Clusters} conducts a study comparing the scaling relations of galaxy clusters in our simulations with observations. Section \ref{Results} delves into our results, covering the identification of clusters in our simulations, their associated significance, and an in-depth analysis of their four primary properties, the mass encompassed within $R_{500}$ ($M_{500}$), the X-ray derived luminosity ($L\rm{x}_{500}$) and temperature ($T\rm{x}_{500}$), and its SZ effect signal
($Y\rm{sz}_{500}$). Finally, in Sect. \ref{Discussion and Conclusions}, we summarize our conclusions and findings.


\section{The SLOW simulations}
\label{The SLOW simulations}
\subsection{Building the initial conditions for SLOW}
Constrained simulations stem from initial density and velocity fields that have been constrained with observational data. The two main pillars on which constrained ICs are constructed are thus the prior cosmological model, like any random simulations, and the observational data. For the former, we use \textit{Planck} cosmology \citep[$\Omega_m$=0.307 ; $\Omega_\Lambda$=0.693 ; $H_0$=67.77 km~s$^{-1}$~Mpc$^{-1}$ and $\sigma_8$=0.829][]{PLANCK2014}. For the latter, in our case, the observational data consist of CF2, the second catalog of the Cosmicflows project \citep[][]{2013AJ....146...86T}. This catalog contains more than 8000 distance moduli of local galaxies obtained from various distance indicators such as the Tully-Fisher relation, the fundamental plane, and supernovae. These distance moduli are combined with observational redshifts to determine galaxy radial peculiar velocities. The method to build the constrained ICs has been extensively discussed in for instance \citet[][]{sorce2016cosmicflows}. The basis for SLOW's ICs have been developed and presented in \citet[][]{Sorce_2018}. They represent CLONES \citep[Constrained LOcal \& Nesting Environment Simulations,][]{Sorce_2021, 2023arXiv230101305S} state-of-the-art so far. We present here a summary of the main steps:

\begin{enumerate}
\item We start with removing any nonlinear motions that would affect our linear reconstruction of the linear initial fields. We are careful to keep galaxies still infalling onto clusters outside of the latter. Galaxies are thus grouped resulting in 5562 galaxies and groups \citep[][]{Sorce_Tempel}.
\item We then minimize the biases inherent to any observational radial peculiar velocity catalogs. The effects of the homogeneous and inhomogeneous Malmquist biases as well as that of the lognormal error \citep[e.g. Kaptney, 1914; Malmquist, 1920;][]{1992ApJ...391..494L, 2016AJ....152...50T} are minimized in the catalog with the method described in \citet[][]{Sorce_2015}.
\item The Wiener-Filter method \citep[WF,][]{Zaroubi_1999} is then applied to the grouped catalog of bias-minimized radial peculiar velocity constraints to reconstruct the tridimensional cosmic displacement field.
\item The reconstructed cosmic displacement field is then used through the Reverse Zel'dovich Approximation \citep[][]{Doumler_2013c} to relocate the galaxy and group constraints to the positions of their progenitors. It ensures that structures are at the proper position at redshift zero. We also replace noisy radial peculiar velocities by their 3D WF reconstructions \citep[][]{2014MNRAS.437.3586S}

\item We produce the initial density and velocity fields constrained with the grouped, bias-minimized, and relocated 3D peculiar velocities using the constrained realizations technique \citep[CR,][]{1991ApJ...380L...5H, 1992ApJ...384..448H, 1996MNRAS.281...84V}. This method derives an estimate of the residual between the WF reconstruction and the true field using a random realization. It permits the regions with poor and noisy data with a random Gaussian field to be filled up while converging to the constrained values where these are available. Doing so we restore statistically the otherwise missing structures and we ensure the overall compliance with the prior power spectrum. The "strength" of the observational data used as constraints determines the extent to which the ICs are likely to reproduce the observed local Universe. In those regions where the data is lacking or dominated by the error, the recovered velocity field will tend to the random realization of the prior model. The outcome of the simulation will thus be determined by the interplay between the random modes and the constraints.
\item As scales below the linear threshold (3 -- 4 Mpc) are non-constrained, 
at least not directly\footnote{It can though be shown that some small scales are induced by the large-scale environment. \citet[][]{Carlesi_2016} showed that it is the case of the local Group.}, there is no reason to work with density and velocity fields with resolution higher than that of the galaxy groups and clusters. The resolution of the constrained density and velocity fields can then be increased by adding small-scale features using the \textsc{Ginnungagap} software\footnote{https://github.com/ginnungagapgroup/ginnungagap}. 
\end{enumerate}

\subsection{The SLOW simulation set}
\label{The SLOW Simulation Set}

Our simulations are based on the parallel cosmological Tree-PM code P-Gadget3 \citep[][]{2005MNRAS.364.1105S}, with which we ran several cold dark matter (DM) and hydrodynamical runs at different resolutions. The code uses an entropy-conserving formulation of SPH \citep[][]{2002MNRAS.333..649S}, with SPH modifications according to \citet[][]{2004ApJ...606L..97D, 2005MNRAS.363...29D} and \citet[][]{10.1093/mnras/stv2443},  and follows the gas using a low-viscosity SPH scheme to properly track turbulence \citep[][]{Dolag_2005b}. Hydrodynamical runs include prescriptions %
 for a vast range of baryonic physics, such as cooling, thermal conduction based on \citet[][]{2004ApJ...606L..97D} at 1/20th of the classical Spitzer value \citep[][]{1962pfig.book.....S}, star formation, supernovae \citep[SNe,][]{Tornatore_2004, Tornatore_2007} and supernova driven winds \citep[][]{2003MNRAS.339..289S}, chemical enrichment from stars from SN type 1a and type 2 and the asymptotic giant branch (AGB) stars, black hole growth and feedback from AGN \citep[][]{2005MNRAS.361..776S, Fabjan_2010,Hirschmann_2014,2015MNRAS.448.1504S}. These runs will in the following be referred to as Full Physics (FP) Runs. Independently from those, we performed another run solely including gas and prescriptions for Cosmic Rays and Magnetohydrodynamics \citep[for more details on this simulation see ][]{boess2023simulating}.
Table \ref{table:1} shows a complete list of all the cosmological boxes available at the current date. Throughout this paper, we rely on the outcomes derived from the SLOW-FP1536$^3$ simulation, knowing that masses and positions for the different structures have only minimal differences between simulations. However, this particular simulation offers the highest resolution available, encompassing galaxy physics down to z = 0, as indicated in Table \ref{table:1}.
\\

\section{Local galaxy clusters}
\label{Local galaxy clusters}

\subsection{A general view}
\label{A General View}

\begin{figure*}
  \centering
  
  \begin{subfigure}{\textwidth} 
    \centering
    \includegraphics[width=1.0\linewidth]{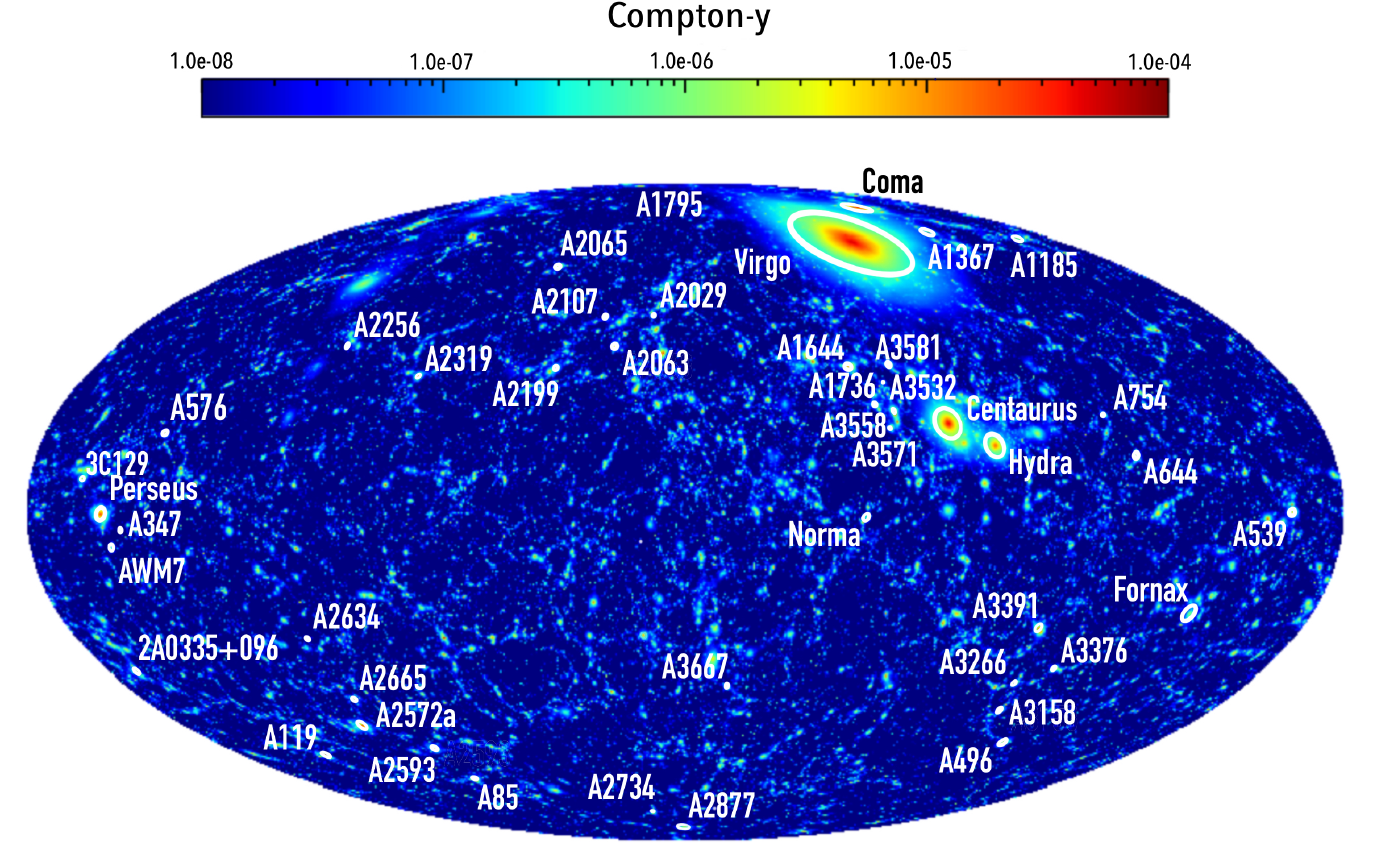}
    \caption*{(a)} 
  \end{subfigure}
    \begin{subfigure}{\textwidth} 
    \centering
    \includegraphics[width=1.0\linewidth]{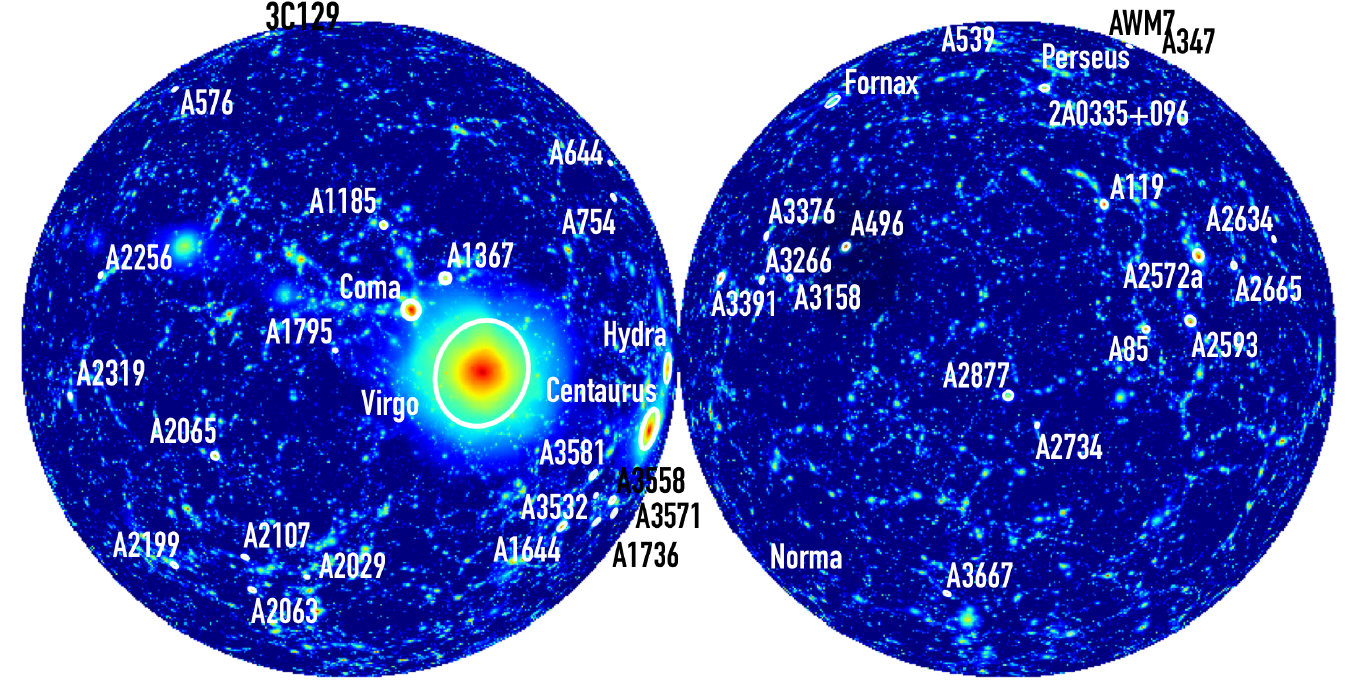}
    \caption*{(b)}
  \end{subfigure}
  \caption{ Full-sky projections of the Compton-y signal encompassing the entire simulation volume in SLOW are shown in the following manner: (a) A full-sky projection in galactic coordinates. (b) On the left, the galactic northern sky is projected, while on the right, the galactic southern sky is shown. The zone of avoidance lies at the edge of the spheres. Circles are employed to mark the projected $r_{500}$ values for the cross-identified clusters. Note that the cross-identification of A576 and A3571 was improved with respect to \citet[][]{Dolag_2023}}.

\label{Full sky plot}
\end{figure*}

The local Universe encompasses a spatial region of approximately z~<~0.1 (R < 200 Mpc/h).
Due to its close proximity,
the local Universe is the most extensively studied part of the Universe, presenting a unique laboratory to study 
galaxy clusters 
across a wide range
of shapes, characteristics, and dynamical states. 

In our local volume, located approximately 15 Mpc away, the Virgo cluster stands out as the primary defining feature of our immediate cosmic neighborhood and covers the largest area in the sky. Its closeness has made it among the best observed clusters of our Universe \citep{2000eaa..bookE1822B}, also hosting the well-known central galaxy M87.

Moving beyond Virgo, a roster of renowned and well-observed clusters such as Centaurus, Fornax, Hydra, Norma, and Perseus take the central stage within the local volume. Perseus, in particular, has been widely observed through X-ray studies using Chandra observations \citep[][]{Weisskopf_2000, Fabian_2011}, as well as  ROSAT PSPC \citep[][]{1981ApJ...248...55B, 1992A&AS...96...23S, 1993MNRAS.262..901A}, Hitomi \citep[][]{2017ApJ...837L..15A}{}{}, XMM-Newton data and Suzaku mosaics \citep[][]{Simionescu_2012, Simionescu_2013, Urban_2013}.
These studies have shown very distinctive features that make this cluster worth studying, like its pronounced cool core, with a sharp peak in X-ray surface brightness and decreasing temperature toward the center, and strong signs of ongoing AGN feedback processes \citep[][]{1993MNRAS.264L..25B, 2000PASP..112.1145F}. Perseus shows also signs of a powerful merger, yet its cool core remains undestroyed. However, the motion induced by this merger seems to penetrate the cool core of the cluster forming a large-scale sloshing inside the cluster \citep[][]{Simionescu_2012}. 

\begin{table} %
\caption{Set of available SLOW simulations following different physical processes and utilizing different resolutions. All are based on the same initial constrained density and velocity field, adding small-scale features up to the $6144^3$ resolution and then degrading resolution to the given value.}

\label{table:1}

\begin{tabular}{|l|c|p{1.2cm}|p{1.4cm}|p{1.2cm}|p{1.2cm}|}

 \hline
 Name & $z$ & N & $M_{\mathrm DM}$    & $M_{\mathrm gas}$   & Physics \\
 SLOW -- &     &   & [$M_{\odot}h^{-1}$] & [$M_{\odot}h^{-1}$] & \\
 \hline
 DM768$^3$  & 0 & 768$^3$  & $2.4\times10^{10}$  & -- & DM only \\
 \hline
 DM1576$^3$ & 0 & 1576$^3$ & $2.9\times10^{9}$ & -- & DM only\\
 \hline
 DM3072$^3$ & 0 & 3072$^3$ & $3.7\times10^{8}$ & -- & DM only\\
 \hline
 DM6144$^3$ & -- & 6144$^3$ & $4.6\times10^{7}$ & -- & ICs only\\
 \hline
 AGN768$^3$  & 0 & 2$\times 768^3$  & $2.0\times10^{10}$ & $3.7\times 10^{9}$  & Cooling, \\
             &   &                  &                    &                     & SFR, AGN\\
 \hline
 AGN1536$^3$ & 0 & 2$\times 1536^3$ & $2.5\times10^{9}$  & $4.6\times 10^{8}$  & Cooling, \\
             &   &                  &                    &                     & SFR, AGN\\
 \hline
 AGN3072$^3$ & 2 & 2$\times 3072^3$ & $3.1\times10^{8}$  & $5.5 \times 10^{7}$ & Cooling, \\
             &   &                  &                    &                     & SFR, AGN\\
 \hline
 CR3072$^3$ & 0 & 2$\times 3072^3$ & $3.1\times10^{8}$ & $5.5 \times 10^{7}$ & MHD, \\
            &   &                  &                   &                     & Cosmic Rays\\
\hline

\end{tabular}

\end{table}

At a distance of roughly 100 Mpc, we find the Coma cluster, a remarkable rich and complex structure, with an also rich history of remarkable observations \citep{ 1987ApJ...317..653F, 1988A&A...199...67M, 1996A&A...311...95B, 1998ucb..proc....1B, 2013A&A...554A.140P}.
\citet[][]{1993Natur.366..429W} showed that the Coma cluster was plausibly formed by the merging of several distinct substructures
which are not yet fully merged
and \citet[][]{1994ApJ...435..162V} showed that the extended regions of X-ray emission in the central region of Coma are associated with the subgroups NGC 4889 and 4874, two galaxies lying at the center of the cluster.  It also presents signs of recent infall in the form of linear filaments to the southeast   \citep[][]{Vikhlinin_1997}, extending $\approx$ 1 Mpc from the cluster center toward the two central galaxies aforementioned \citep[][]{Andrade_Santos_2013, 2001A&A...365L..74N, 2001A&A...365L..67A, 2001A&A...365L..60B}. 

Beyond the previously mentioned clusters, we can find a significant number of very massive clusters extending up to a distance of over 150 $h^{-1}$Mpc \citep[][]{ 2013AJ....146...69C, refId0}. Therefore, within the cosmological volume where our simulation is constrained, we account for a rich variety of galaxy clusters. This allows us to conduct detailed studies of their evolution as well as testing (hydro)dynamical and physical processes governing their formation.

\subsection{Collection sample of local galaxy clusters}

The previously mentioned examples show the immense diversity of processes and dynamical states observed in local clusters. Thus, detailed observations of the local Universe add immeasurable value to astrophysical discussions like the survivability
of cool cores \citep[e.g.][]{2013ApJ...774...23M}, the effects of mergers and AGN associated feedback \citep[e.g.][]{2021Univ....7..142E}, as well as cosmological discussions on large scale structure formation \citep[e.g.][]{2002MNRAS.333..739M}, mass estimation based on individual clusters \citep[e.g.][]{2023arXiv231002326L} and inference of cosmological parameters \citep[e.g.][]{2023ApJ...954..169T}.%

In this work, we collected information from a total of 221 local Universe clusters and groups from the literature. To do so, we combined the {\it CLASSIX} catalog of local X-ray clusters and groups \citep{2016A&A...596A..22B}, the Tully galaxy groups catalog\footnote{https://edd.ifa.hawaii.edu} \citep[][]{2015AJ....149..171T} and the SZ Cluster Database\footnote{http://szcluster-db.ias.u-psud.fr}. This list was then augmented with additional data from the X-rays Galaxy Clusters Database (BAX)\footnote{http://bax.irap.omp.eu}.
In addition, we included other, well-known local clusters and groups from various, individual observations. From this collection, we selected all clusters with SZ-effect derived $M_{500}$ masses\footnote{$M_{500}$ is the mass within the radius $R_{500}$ for which the cluster mean total density is 500 times the critical density at the cluster redshift} above $2\times10^{14}$ M$_{\odot}$, with a corresponding position in the Tully galaxy group catalog. Additionally, it incorporated six renowned local clusters with an X-ray-derived $M_{500}$ value exceeding $10^{14}$~M$_{\odot}$ as well as prominent galaxy clusters from the Tully galaxy groups catalog where additional X-ray data where available. From this, we constructed a final set of 45
clusters where we identified a counterpart candidate within the simulations. Within this selection, the inferred masses of the clusters are based on vastly different methods. Therefore, in the later analysis, we restricted direct comparisons of $M_{500}$ to the subsample of clusters where $M_{500}$ was based on the observed SZ signal.

\subsection{Finding cluster replicas in SLOW}

For the sample of 46 galaxy clusters, we collected data from the literature for different mass estimates and X-ray observables. The integrated Compton-y signal (Y) within $R_{500}$ together with the corresponding $M_{500}$ mass estimate was taken from the SZ Cluster Database, while X-ray data like temperature and luminosity in the 0.1-2.4 keV range of most of the clusters where taken directly mainly from \citep[][]{Ikebe_2002, Shang_2008, 2011A&A...536A..11P}, otherwise we took the values quoted in BAX. The inferred X-ray based $M_{500}$ masses where extracted from \citep[][]{Chen_2007}. The Tully galaxy groups catalog presented dynamical mass estimations, $M_{\rm{dyn}}$, which we converted to the virial mass, by correcting it down by 12\% as presented by \citet[][]{Sorce_2016} for their simulated Virgo cluster at z = 0, and then converted the virial mass to $M_{500}$ using the conversions presented in \citet[][]{2021MNRAS.500.5056R}.

In Table \ref{table:2}, we report the observed position as sourced from the according group within the Tully galaxy catalog, together with the position of the replica candidate within our simulation. In addition, we list the relative displacement between the observed cluster position and that of the replica candidate. Table \ref{table:3} displays the observational values alongside the corresponding values from the replica candidate, as obtained through the simulation, for comparison.
This now allows a more thorough comparison and emphasizes the strength of comparing multiple observational signals to the counterparts from the simulation. Typically, X-ray luminosity is among the most commonly measured quantities, %
although it is most sensitive to inhomogeneities such as clumping and hence most dependent on the detailed treatment of cooling and star-formation processes in simulations. Therefore, it is difficult to accurately predict through simulations. In a fully virialized cluster formed solely through gravitational collapse, X-ray luminosity, temperature as well as SZ signal directly correlate with the mass of the cluster. However, in practice, such relations suffer differently from scatter due to internal structures, deviation from spherical symmetries as well as deviation from hydrostatic equilibrium. Indeed, different comparisons suffer differently from individual observational biases, as well as from the incompleteness of the simulations when it comes to reproducing the actual galaxy clusters and the treatment of relevant physical processes affecting them. Therefore, we can get a more complete picture if we make use of the entire set of observables along with the inferred masses for comparison.

In order to obtain the above-mentioned sample of counterpart candidates  
of local galaxy clusters within the SLOW simulation, we applied the following procedure:

\begin{enumerate}
\item  Starting from our collection of observed local galaxy clusters we first associated the cluster/group in Tully's North/South Catalog. We used the member galaxy which is closest to the position on the sky and redshift as the center and used the according supergalactic X, Y, Z coordinates. For the few observed clusters without such an association, we converted sky positions and redshifts directly into supergalactic X, Y, Z coordinates. When computing the positions of clusters within the simulation, we used the observer position as presented in \citep[][]{Dolag_2023} as the optimized center.

\item We selected the Compton-y derived $M_{500}$ cluster mass, or if unavailable, the X-ray derived mass or dynamical mass, depending on data availability.

\item We then identified all massive halos within a small "search radius," and with a lower masses $M_{500}$ cut typically not less than six times the observed mass. If we fail to identify a suitable counterpart in close proximity to the observed cluster, we expand our search radius typically not exceeding 45 Mpc/h. The resulting median distance in which we found our cluster replicas is $\approx$ 25 Mpc/h.

\item In cases where multiple candidates existed in terms of position and mass, we conducted a comparison of temperature and luminosity values, along with an assessment of the cluster's close surroundings and dynamical state, especially if it was involved in a merging process, or in a stage prior to it. Additionally, for supercluster regions like Hydra-Centaurus or Shapley, 
we considered the geometry of the surrounding cluster environment, its relative position to nearby clusters, and how well the shape of the surrounding filaments matched the observed galaxy distribution (see also Seidel et al., in prep).

\end{enumerate}

Although the selection process described above is guided by the availability of observational data, the combined usage of \textit{Planck} selected clusters, the X-ray flux-limited {\it CLASSIX} catalog and the 2MRS-based Tully galaxy groups catalog should lead to a very complete cover of massive local galaxy clusters. %
Such a list of counterpart candidates is specifically designed to focus primarily on the massive clusters identified through observational proxies. A comprehensive map of the entire sky, featuring the 46 cluster replicas included in our SLOW simulation set, is provided in Figure \ref{Full sky plot}. We note that this sample of clusters not only contains the 13 most massive clusters but also 70\% of clusters with $M_{500}$ larger than $2\times10^{14}$ M$_\odot$ from the \textit{Planck} SZ cluster catalog \citep[][]{2014A&A...571A..29P}{}{} 
within a distance of $\approx200$ Mpc/h.

\section{Scaling relations for low redshift clusters}
\label{Scaling Relations for Low Redshift Clusters}

\begin{figure*}
\begin{multicols}{2}

    \includegraphics[width=\linewidth]{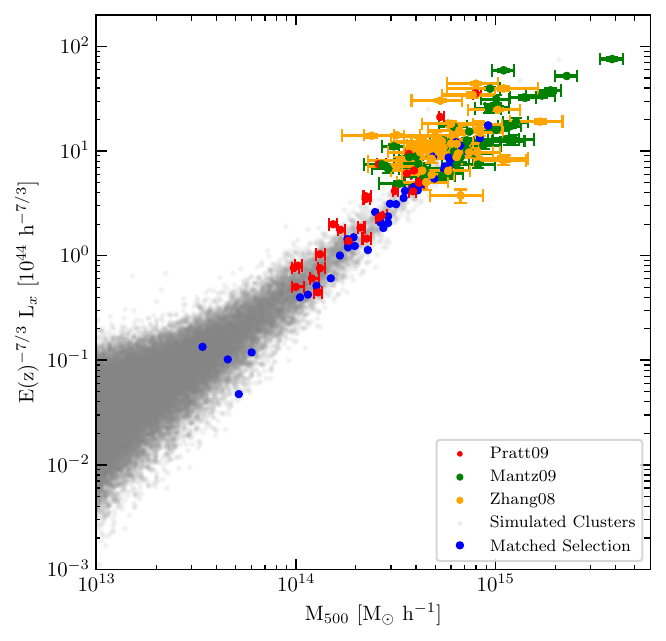}\par 
    \includegraphics[width=\linewidth]{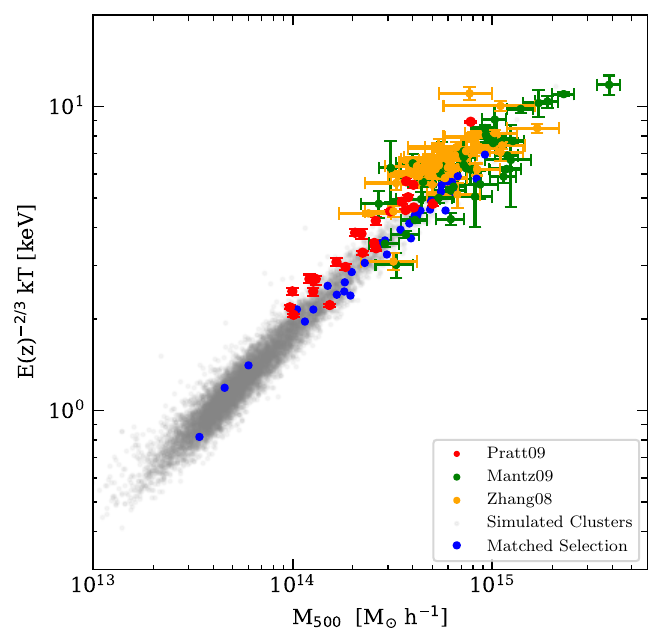}\par
    \includegraphics[width=\linewidth]{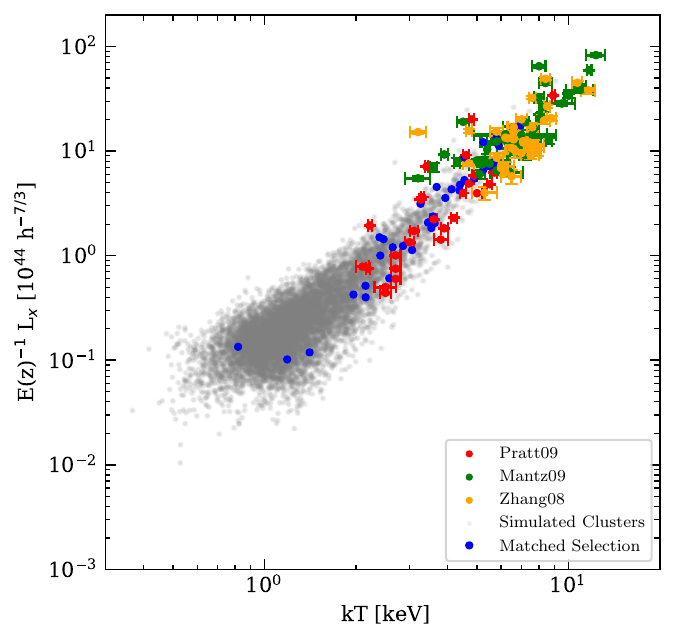}\par 
    \includegraphics[width=\linewidth]{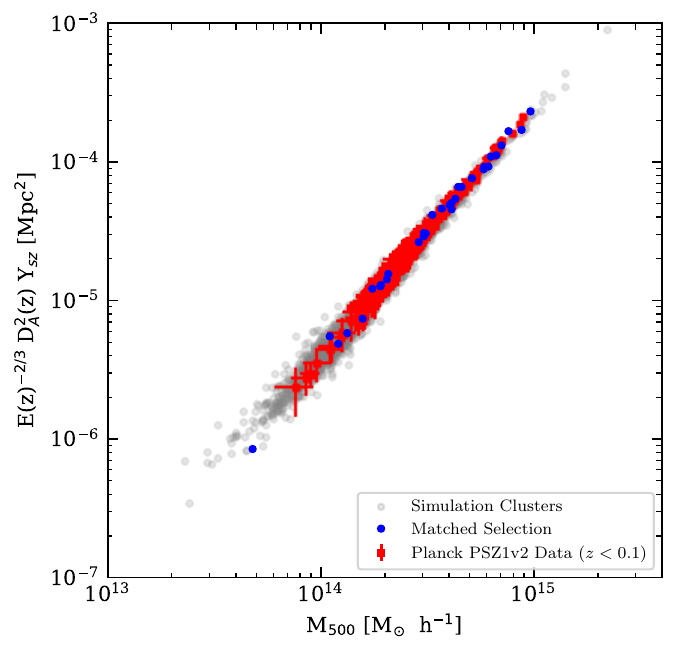}\par

\end{multicols}
\caption{Scaling relations within $R_{500}$ for simulated clusters at z = 0 and for observed low redshift clusters. Gray data points represent the cluster values from our simulations, estimated using \textsc{Subfind} \citep[][]{2001MNRAS.328..726S, Dolag_2009, Saro_2006}. The clusters that have been cross-identified are highlighted in blue. Luminosity-mass ($L_{\rm{x}}$ -- $M_{500}$, top left), luminosity-temperature ($L_{\rm{x}}$ -- kT, top right), temperature-mass (kT -- $M_{500}$, bottom left) panels show observational data from \citet[][]{2009A&A...498..361P}{}{} (depicted as red points), which encompasses clusters with redshifts ranging from 0.05 to 0.164, \citet[][]{2008A&A...482..451Z}{}{} (depicted in yellow) with clusters spanning the redshift range of 0.14 to 0.3, and \citet[][]{2010MNRAS.406.1773M}{}{} (depicted in green), which includes clusters with redshifts less than 0.3. The bottom right panel depicts the Compton-y -- $M_{500}$ scaling relation. For observational data in this case, we rely on Planck's catalog PSZ1v2 \citep[][]{2014A&A...571A..29P}{}{},
considering only clusters with a redshift of less than 0.1.}%
\label{scaling relations}
\end{figure*}

It has been extensively demonstrated that the subgrid model presented in Sect. \ref{The SLOW Simulation Set} yields galaxy and ICM properties in galaxy clusters that closely align with observed trends and properties \citep[][]{2017MNRAS.469.3069G, 2020MNRAS.494.3728S}{}{}. The Magneticum simulations %
employing the same
hydrodynamical scheme, with a slightly earlier adaption of the galaxy formation treatment, have been compared to SZ data from \textit{Planck} \citep[][]{PLANCK2014} and SPT \citep[][]{McDonald_2014}. They have successfully replicated the observable X-ray luminosity relation \citep[][]{2013MNRAS.428.1395B}, among other ICM characteristics. In this context, we will test the results of our SLOW simulations %
to demonstrate their capability to reproduce SZ data and relationships with X-ray observables scaling relations. 

Scaling relations are particularly interesting as they are tightly related to the physics of cluster formation and evolution. In the idealized
framework where gravity is the dominant process in cluster evolution, self-similar models predict simple scaling relations between cluster properties, such as temperature, luminosity, and Compton-y value with total mass \citep[][]{1986MNRAS.222..323K}. In general, these relations are described by power laws, around which some data points scatter according to a log-normal distribution. These relations describe positive correlations, so that largest systems have on average higher values of the correlated parameter. On top of this, as these relations come from the premise of self-similarity and thus gravity domination, any deviation of the relation may be a sign of important hydrodynamical processes taking center stage in the evolution history of these structures. Therefore, scaling relations are an important tool in cosmology as well as in the study of the thermodynamical history of clusters and their ICM.

\begin{figure}
  \centering
  
  \begin{subfigure}{\textwidth} 
    \includegraphics[width=0.45\linewidth]{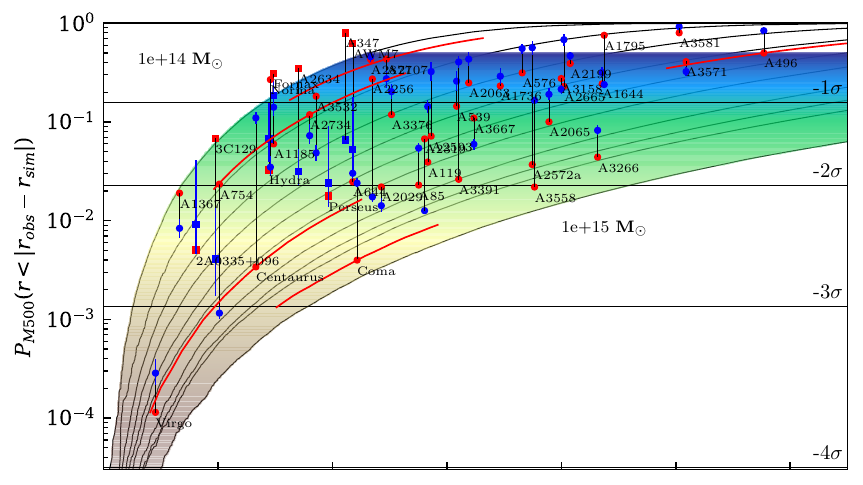}
  \end{subfigure}
    \begin{subfigure}{\textwidth} 
    \includegraphics[width=0.45\linewidth]{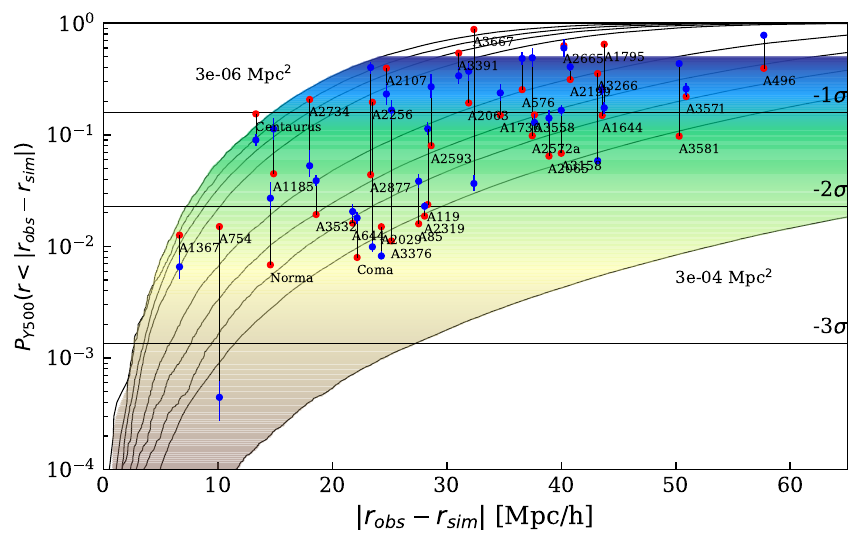}
  \end{subfigure}
  
  \caption{ Cumulative distribution functions for $M_{500}$ and $Y_{500}$, respectively. The top panel displays bins ranging from $10^{15}$ to $10^{14}$ M$_{\odot}$, with increments of ten, showing the random expectation of finding a cluster of a certain mass or higher within a sphere whose radius corresponds to the distance displayed on the x-axis. Simulated $M_{500}$ values, as estimated by \textsc{Subfind}, are represented in red. The blue data points represent the observed $M_{500}$ values obtained from clusters Planck signals, including their associated errors. In cases where SZ-derived $M_{500}$ values were unavailable, X-ray-derived $M_{500}$ values were used (indicated by square points). For cluster A347 X-ray-derived mass information was not available, thus we used the $M_{500}$ mass derived by converting the dynamical mass estimated by Tully. The observational uncertainties in the cluster's position for Coma, A2734, AWM7, and Centaurus are represented by red lines, as an illustrative example of how such positional uncertainty can impact the results of the significance study (see Sect. \ref{Detection Significance Subsection} for a more comprehensive discussion). The bottom panel displays the cumulative distribution function for the SZ-derived signal in $R_{500}$. Simulated datapoint values where estimated using \textsc{Smac} \citep[][]{2005MNRAS.363...29D}. The bins range from $3\times 10^{-6}$ to $3\times 10^{-4}$ Mpc$^2$, with increments of 2.} %
  \label{significance1}
\end{figure}

\begin{figure}
  \centering
  
  \begin{subfigure}{\textwidth} %
    \includegraphics[width=0.45\linewidth]{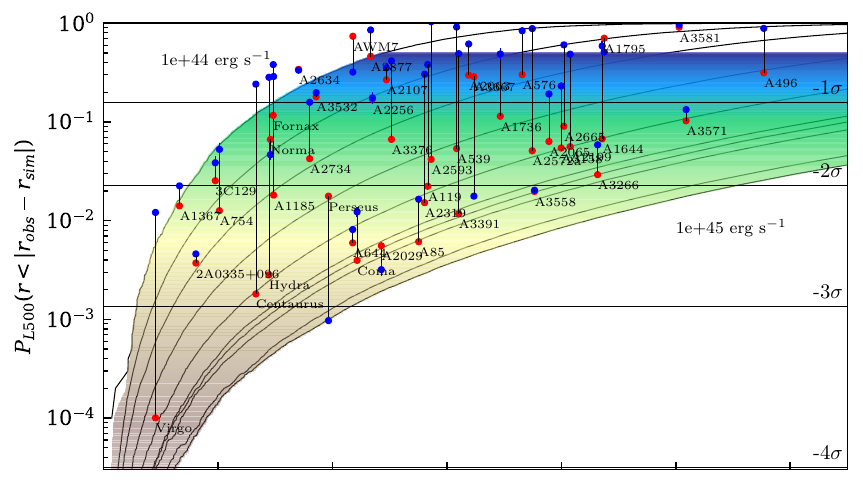}
  \end{subfigure}
    \begin{subfigure}{\textwidth} %
    \includegraphics[width=0.45\linewidth]{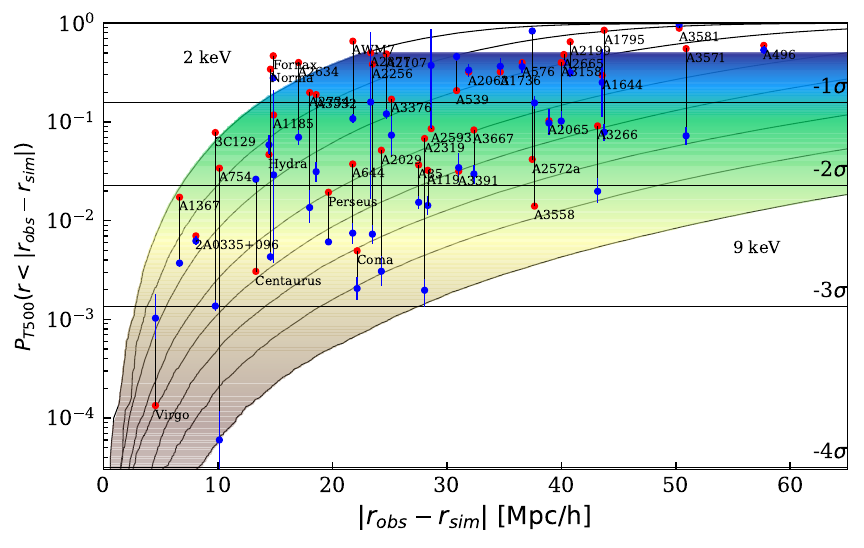}
  \end{subfigure}

  \caption{Similar to Figure \ref{significance1}, we show cumulative distribution functions for $L_{\rm{x}}$ and $T_{\rm{x}}$ within $R_{500}$. For a better comparison with observations, we estimated the X-ray luminosity in the band 0.1-2.4 keV for every cluster using \textsc{Smac}. In the top panel, the bin values range from $10^{44} \rm{erg} \times s^{-1}$ to $10^{45} \rm{erg} \times s^{-1}$, with increments of 10. Uneven spacing between higher-value bins has to do with the inherent limitations of small number statistics within our cosmological box (see Sect. \ref{Detection Significance Subsection} for a deeper discussion). The bottom panel presents the cumulative distribution function from the X-ray temperature in $R_{500}$. We estimated this temperature using \textsc{Subfind}.  The bin values range from 2 keV to 9 keV, with increments of 1 keV.
  } %
  \label{significance2}
\end{figure}

\begin{figure*}
  \centering
  
  \begin{subfigure}{\textwidth} %
    \centering
    \includegraphics[width=1.0\linewidth]{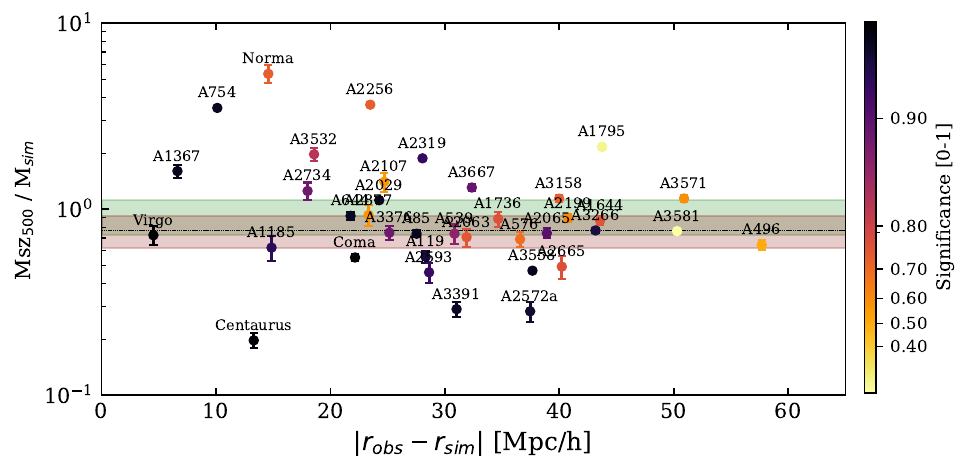}
    \caption*{(a)} %
  \end{subfigure}
  
  \begin{subfigure}{0.33\textwidth} %
    \centering
    \includegraphics[width=\linewidth]{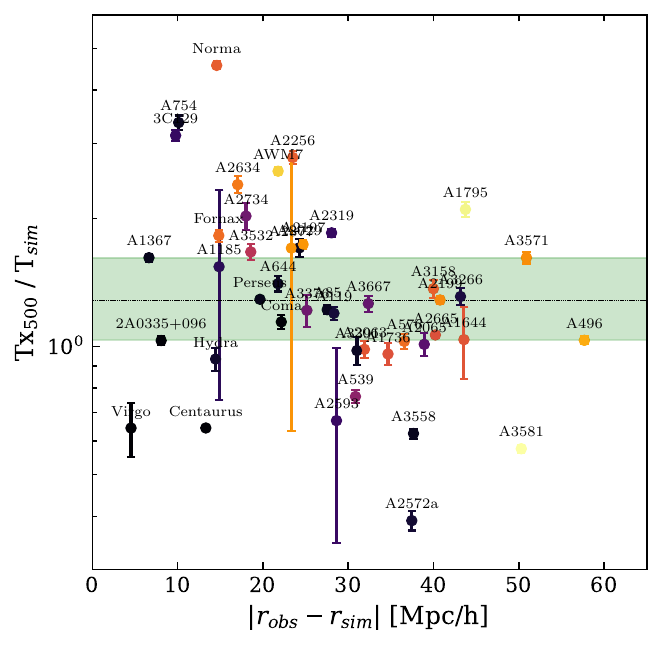}
    \caption*{(b)}
  \end{subfigure}%
  \begin{subfigure}{0.33\textwidth}
    \centering
    \includegraphics[width=\linewidth]{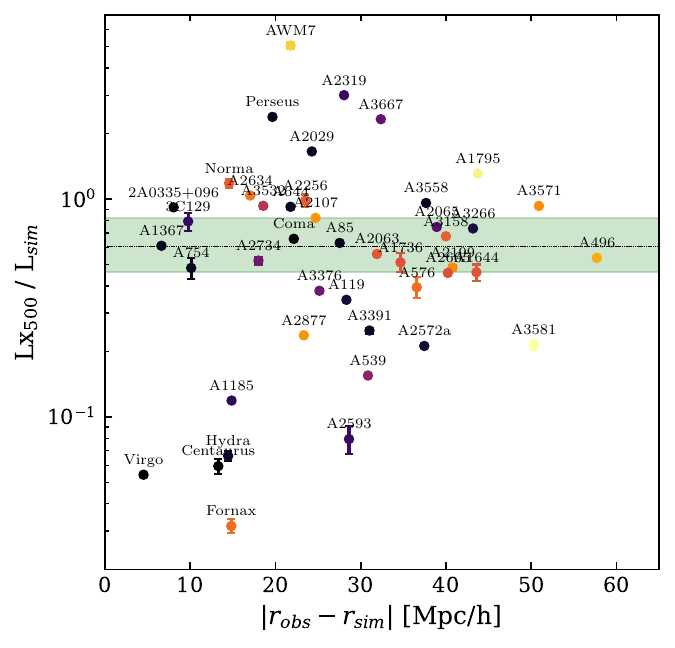}
    \caption*{(c)}
  \end{subfigure}%
  \begin{subfigure}{0.33\textwidth}
    \centering
    \includegraphics[width=\linewidth]{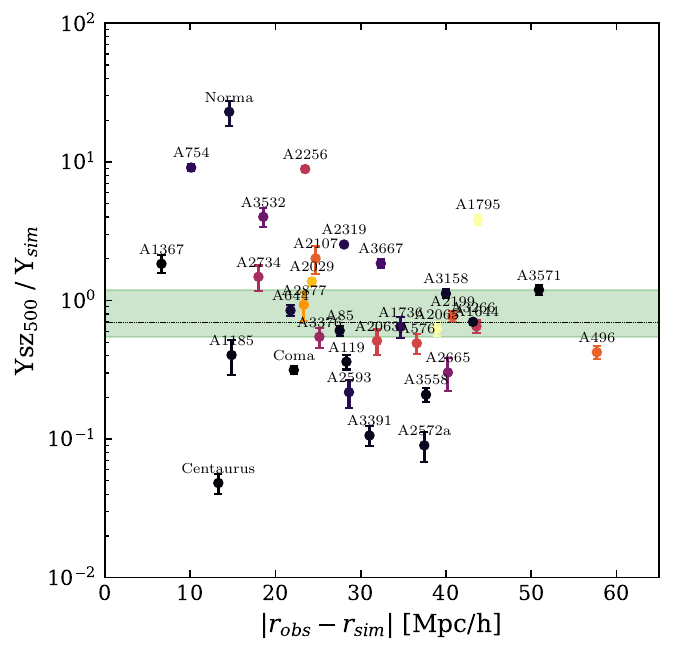}
    \caption*{(d)}
  \end{subfigure}
  
  \caption{Simulated and observed quantities' ratios as a function of relative distance. (a) Ratio between the $M_{500}$ mass, derived from the cluster SZ signal, and the \textsc{Subfind}-estimated $M_{500}$ as a function of relative distance. The dotted line corresponds to the median of the distribution, while the shaded red region depicts the observed dispersion in the Compton-y--mass scaling relations \citep[][]{2011A&A...536A..11P}. The three bottom panels present similar ratios for (b) X-ray luminosity ($L_{500}$), (c) X-ray temperature ($T_{500}$), (d) and SZ-signal ($Y_{500}$). These data points are color-coded based on their mass significance, with higher significance levels indicated in black and lower significance levels denoted in yellow.} %
  \label{property-distance}
\end{figure*}

\begin{figure*}
  \centering
  
  \begin{subfigure}{0.33\textwidth} %
    \centering
    \includegraphics[width=\linewidth]{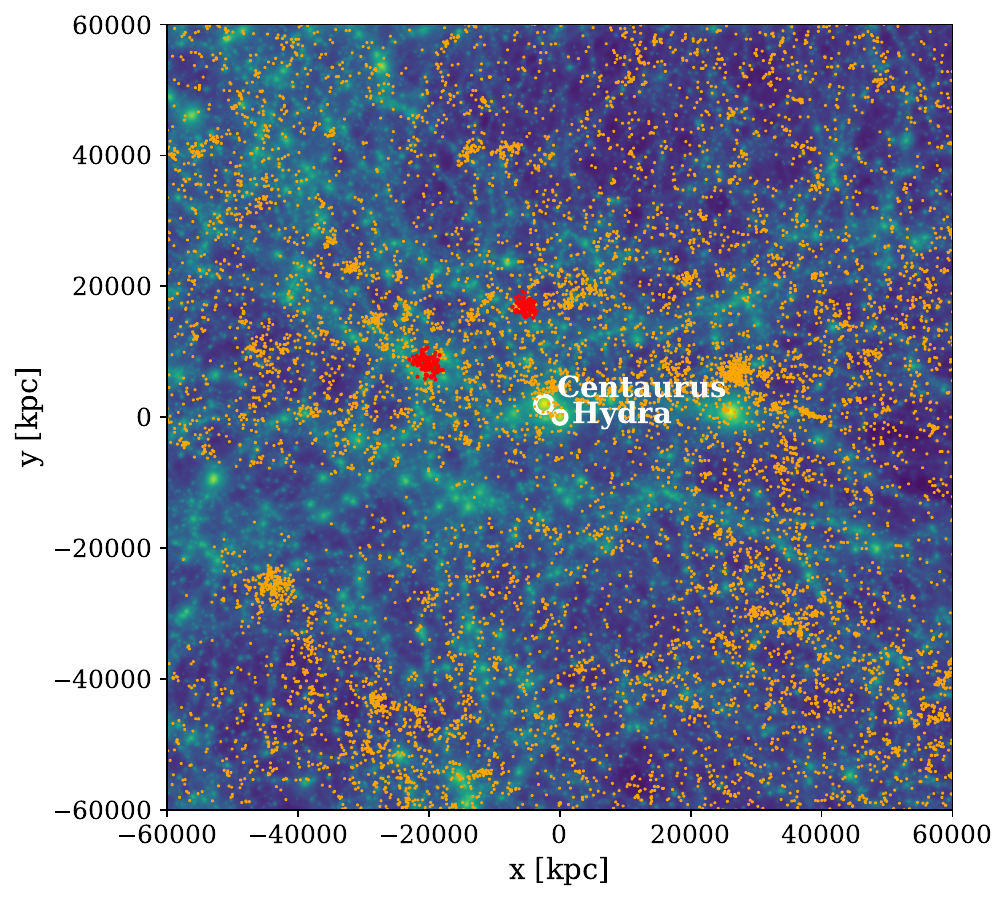}
  \end{subfigure}%
  \begin{subfigure}{0.33\textwidth}
    \centering
    \includegraphics[width=\linewidth]{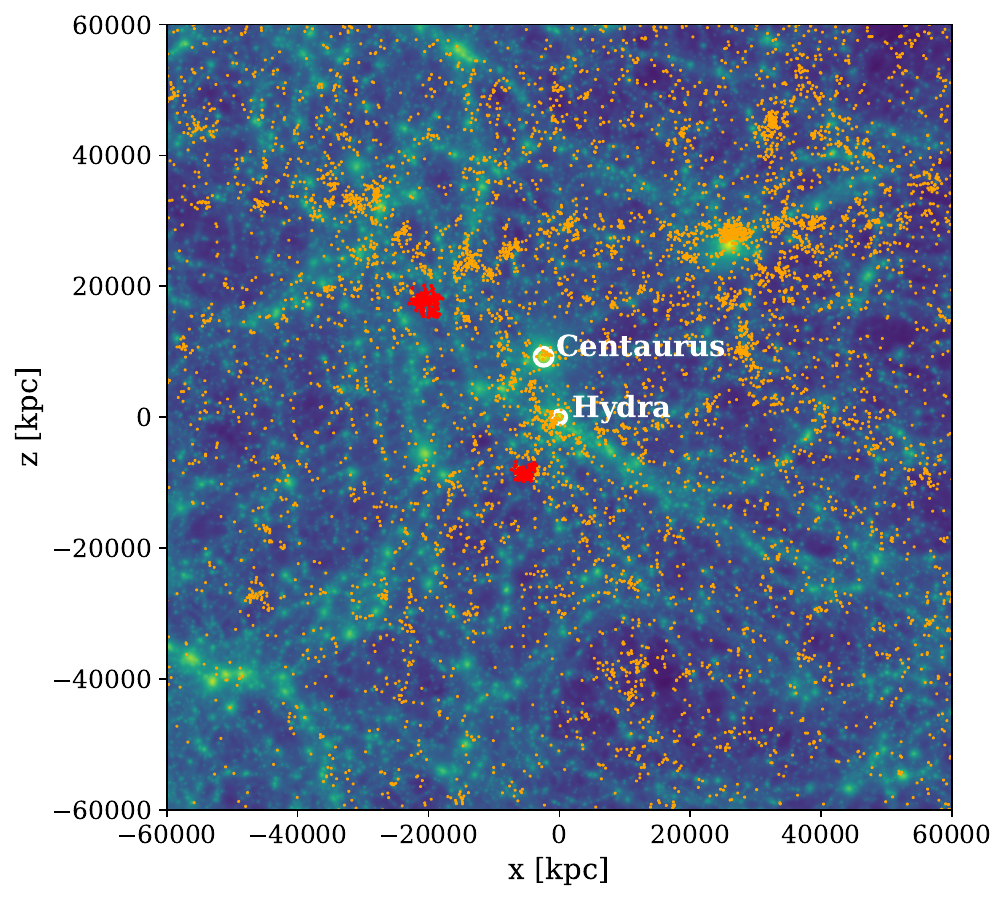}
  \end{subfigure}%
  \begin{subfigure}{0.36\textwidth}
    \centering
    \includegraphics[width=\linewidth]{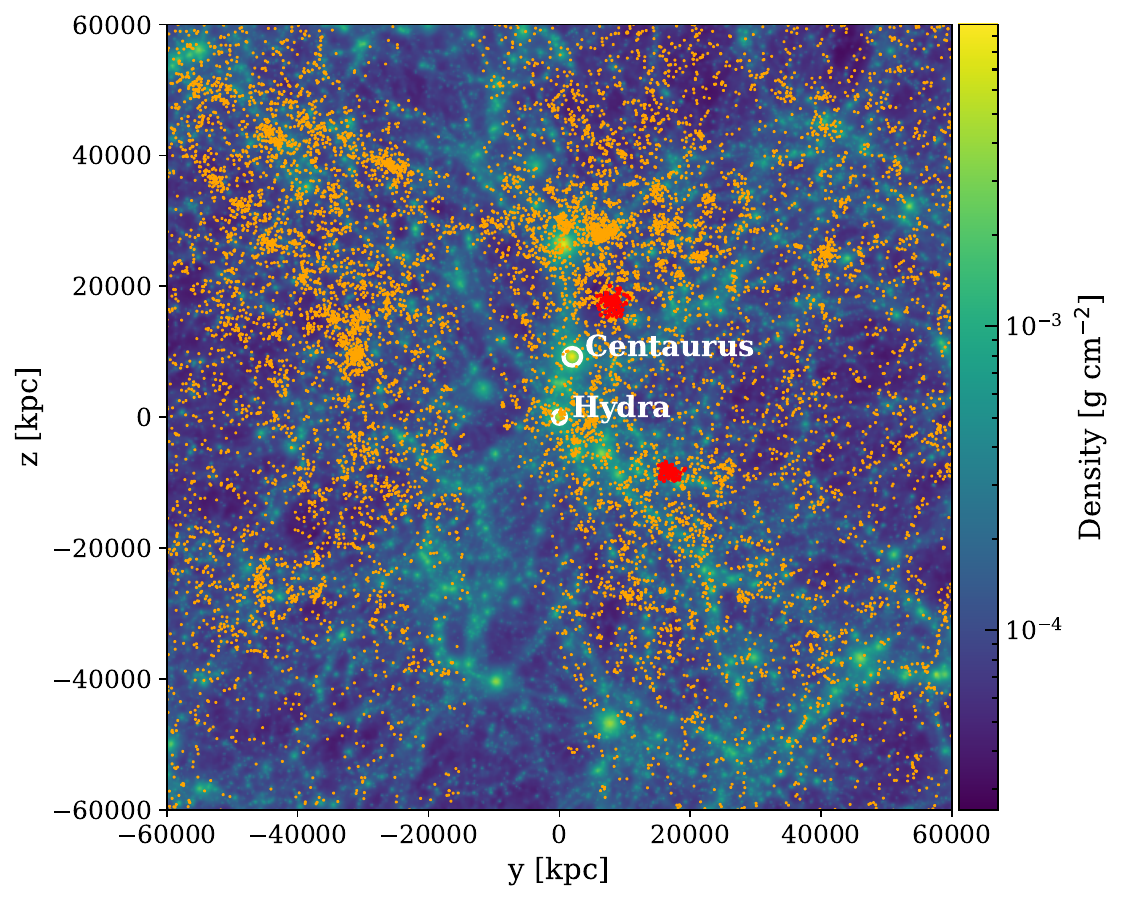}
  \end{subfigure}
  \begin{subfigure}{0.33\textwidth} 
    \centering
    \includegraphics[width=\linewidth]{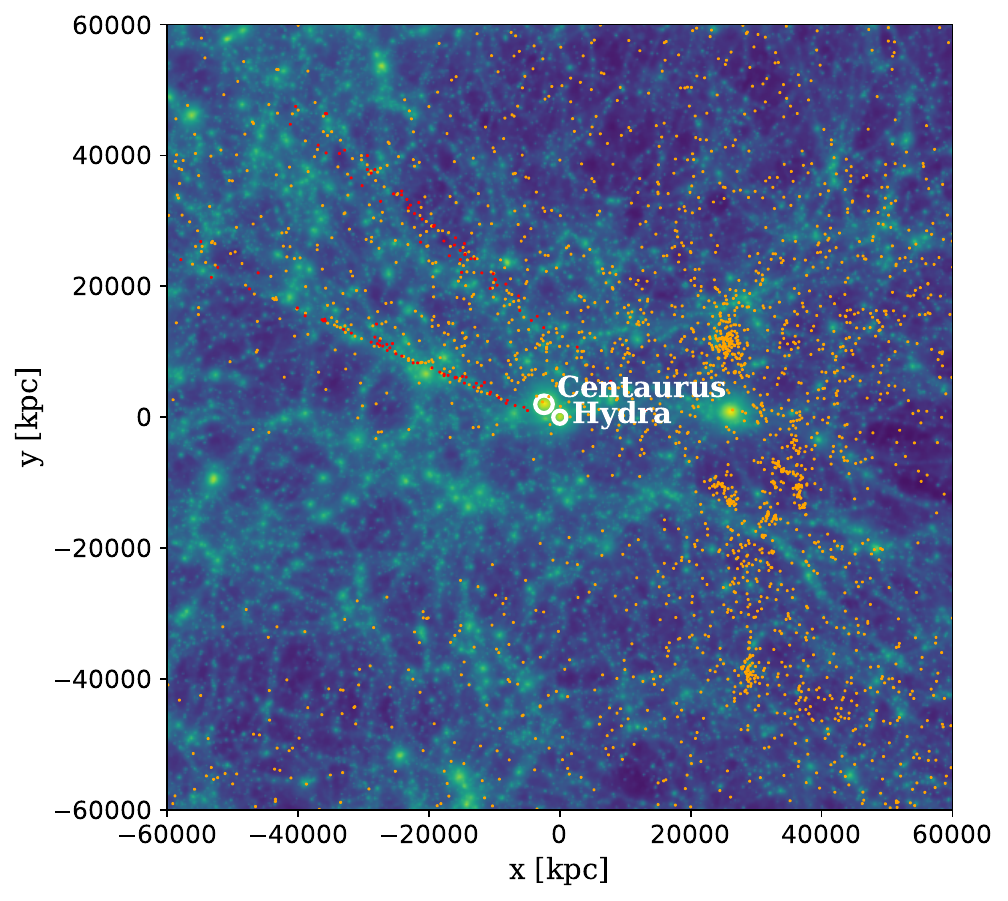}
  \end{subfigure}%
  \begin{subfigure}{0.33\textwidth}
    \centering
    \includegraphics[width=\linewidth]{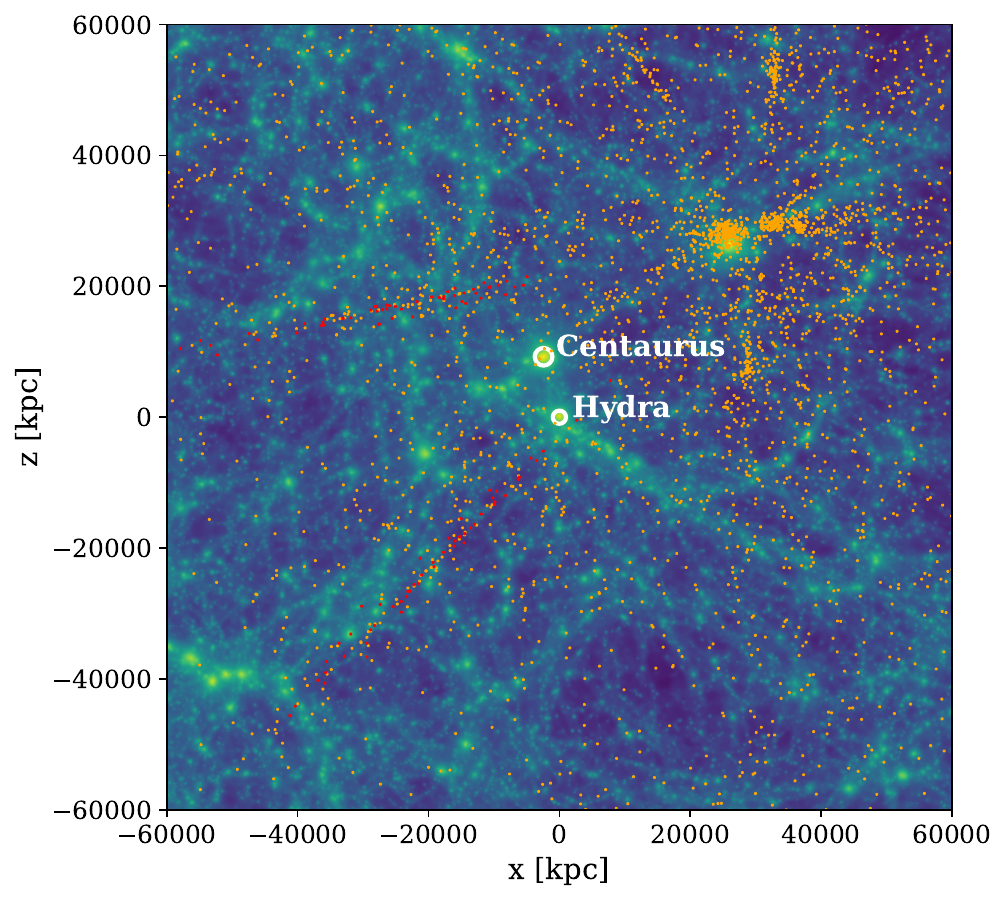}
  \end{subfigure}%
  \begin{subfigure}{0.36\textwidth}
    \centering
    \includegraphics[width=\linewidth]{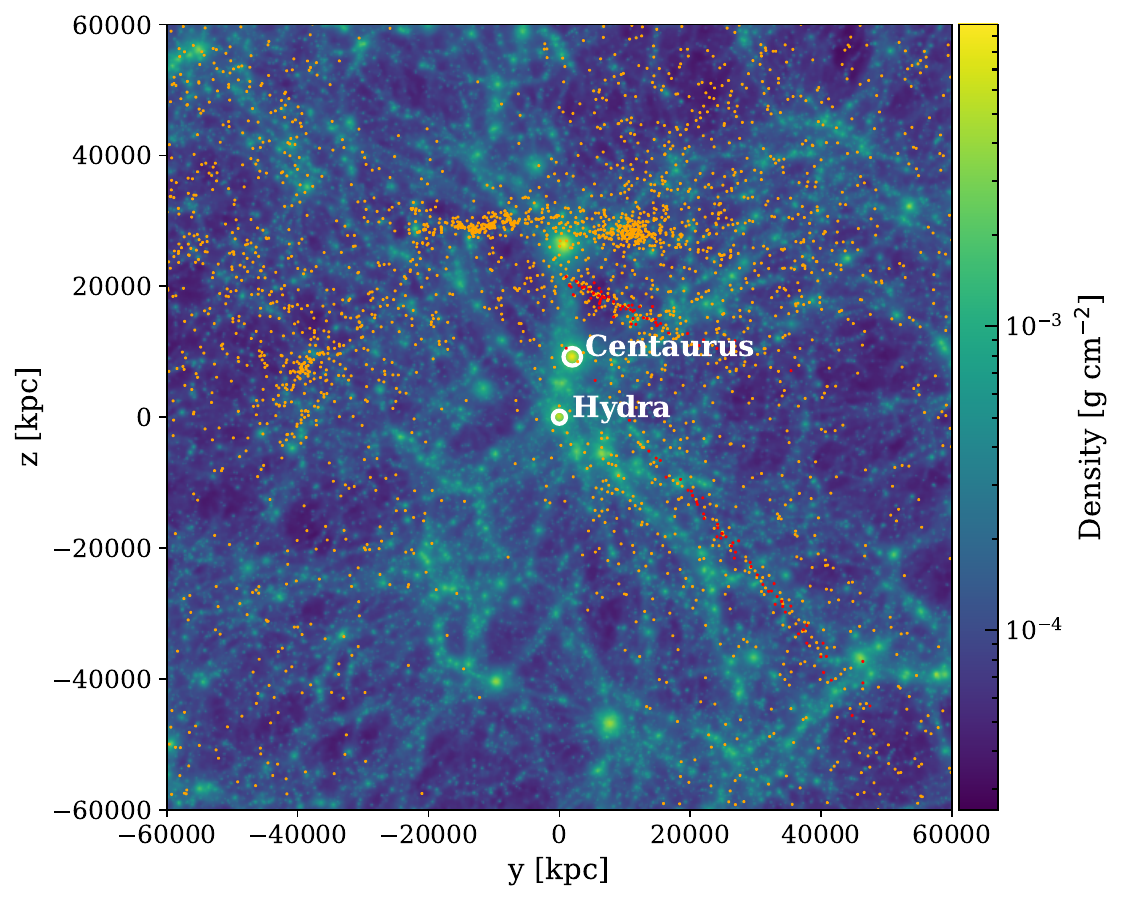}
  \end{subfigure}

  \caption{ Density map of the Centaurus-Hydra region in the SLOW simulation, presented in supergalactic coordinates and created using \textsc{Smac}. The locations of the Centaurus and Hydra clusters are marked with two white circles, each indicating the R$_{500}$ radius of the respective structures. In the top panels, observed galaxies from the 2MRS1175 North Groups catalog associated with Centaurus and Hydra are highlighted in red. Additionally, galaxies listed in both the 2MRS1175 North and South Groups catalogs are shown in orange. These galaxy positions have been corrected to account for distortions such as the "fingers of god" effect. In the bottom panels, all galaxies from the CF-2 Catalog in this area are displayed in orange, with those attributed to Centaurus and Hydra in red, without adjusting for the "fingers of god" correction.}

  \label{Centaurus-Hydra maps}
\end{figure*}

\begin{figure*}
  \centering
  
  \begin{subfigure}{0.33\textwidth} %
    \centering
    \includegraphics[width=\linewidth]{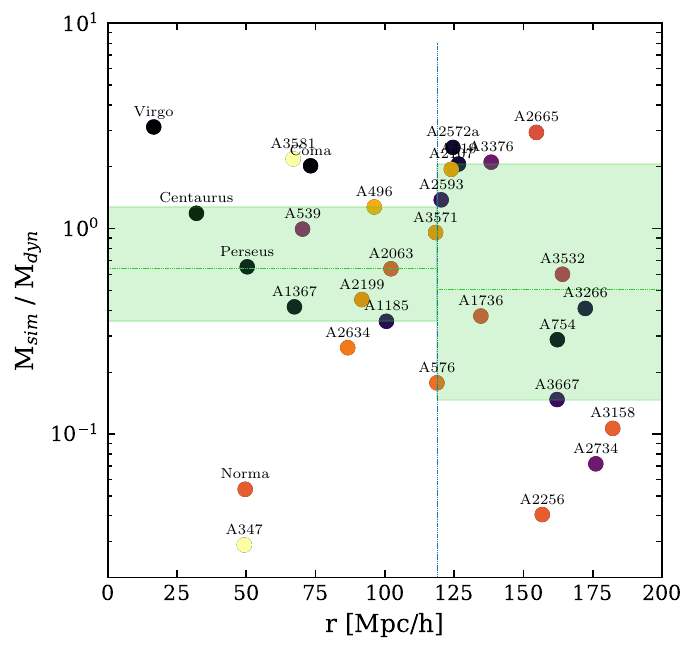}
    \caption*{(a)}
  \end{subfigure}%
  \begin{subfigure}{0.33\textwidth}
    \centering
    \includegraphics[width=\linewidth]{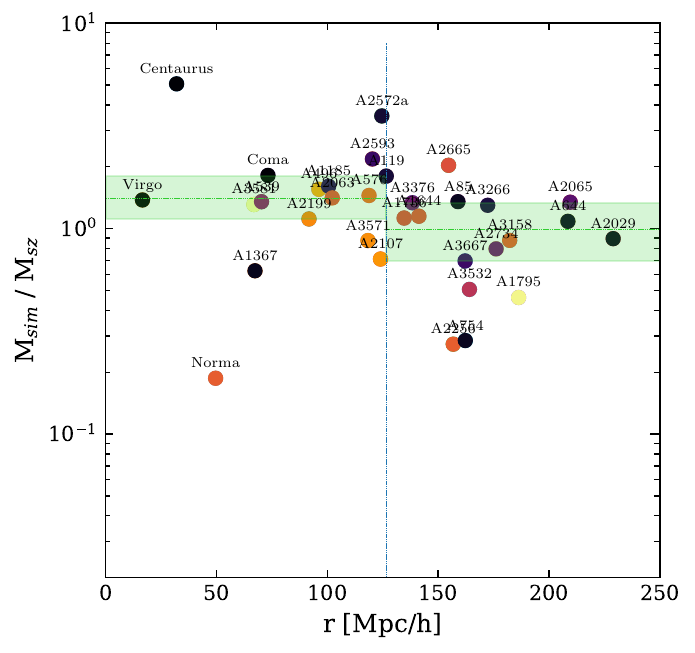}
    \caption*{(b)}
  \end{subfigure}%
  \begin{subfigure}{0.37\textwidth}
    \centering
    \includegraphics[width=\linewidth]{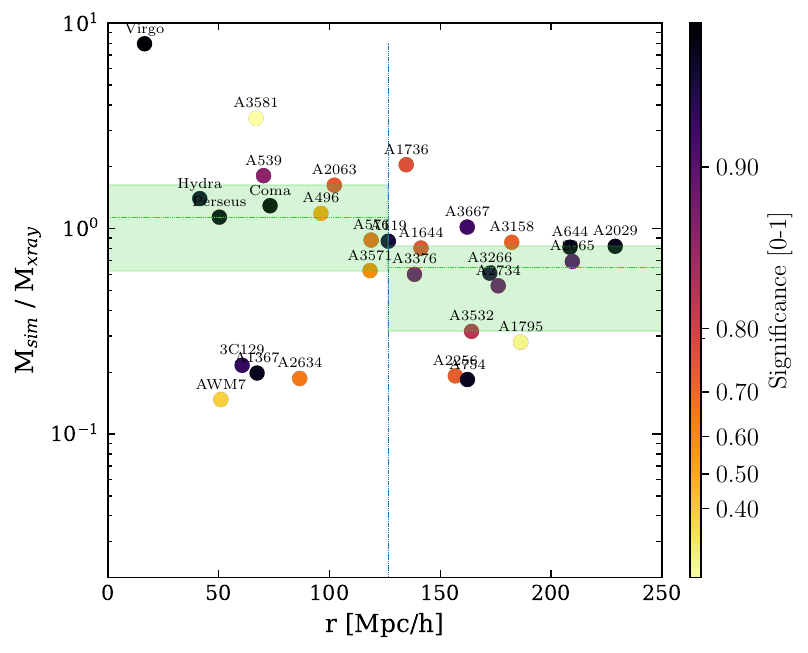}
    \caption*{(c)}
  \end{subfigure}
    \begin{subfigure}{0.33\textwidth} %
    \centering
    \includegraphics[width=\linewidth]{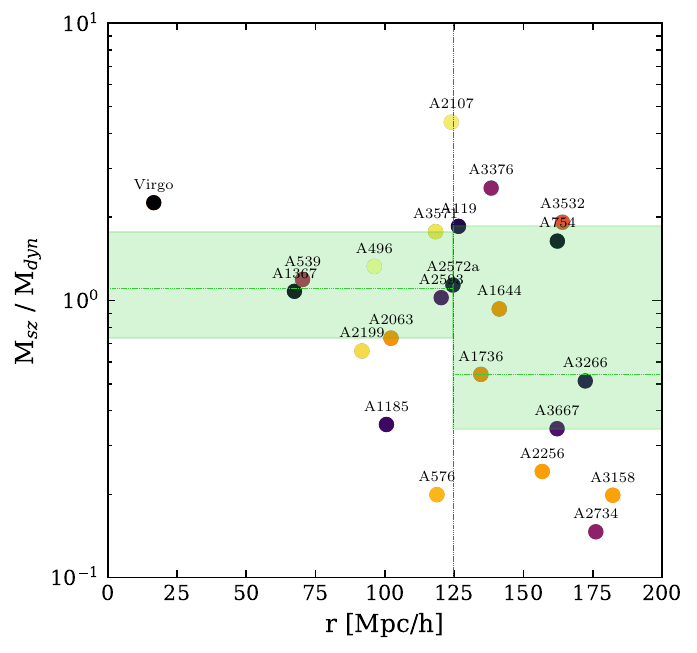}
    \caption*{(d)}
  \end{subfigure}%
  \begin{subfigure}{0.33\textwidth}
    \centering
    \includegraphics[width=\linewidth]{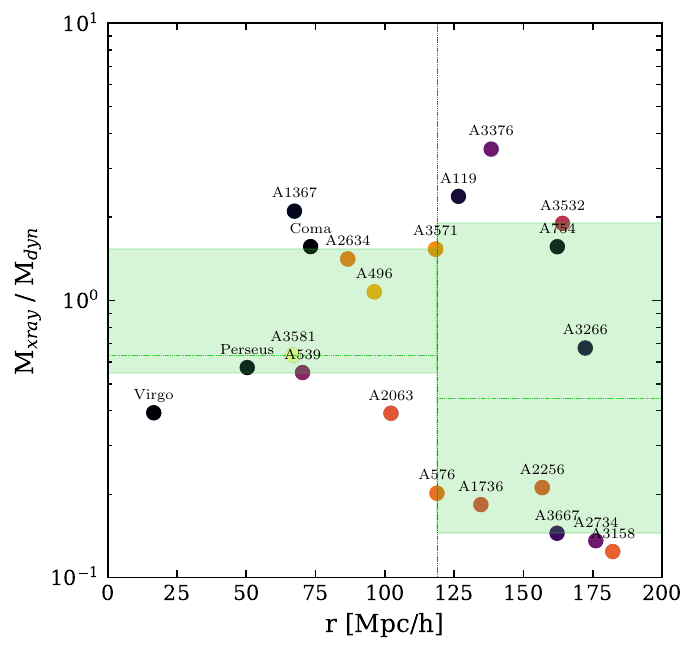}
    \caption*{(e)}
  \end{subfigure}%
  \begin{subfigure}{0.368\textwidth}
    \centering
    \includegraphics[width=\linewidth]{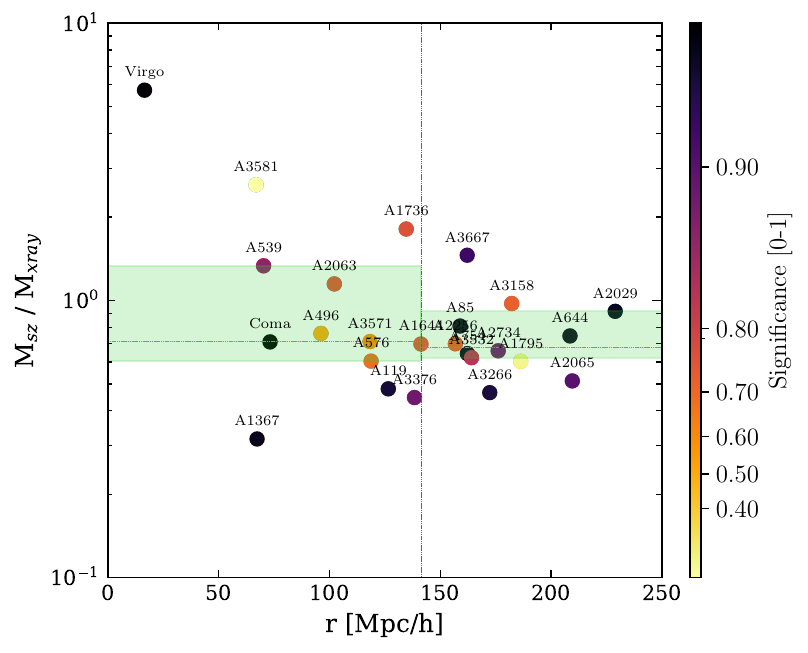}
    \caption*{(f)}
  \end{subfigure}
  
  \caption{ Mass ratios as a function of total distance. The top figures present the mass ratio between \textsc{Subfind}-estimated $M_{500}$ and three different mass estimations: (a) $M_{500}$ derived from the dynamical mass as provided in the Tully galaxy group catalog \citep[][]{2015AJ....149..171T}: $M_{\rm{dyn}}$, (b) $M_{500}$ derived from SZ measurements as available in the Planck database: $M_{\rm{SZ}}$, and (c) $M_{500}$ derived from X-ray observations: $M_{\rm{xray}}$. These ratios are plotted against the total distance of the cluster. In addition, the bottom figures illustrate the ratios between various observational mass estimations: (d) $M_{\rm{sz}}/M_{\rm{dyn}}$, (e) $M_{\rm{xray}}/M_{\rm{dyn}}$, and (f) $M_{\rm{sz}}/M_{\rm{xray}}$, all in relation to the total distance of the cluster. (For further details on the observed quantities, please refer to table \ref{table:3}). A vertical dashed line divides the clusters into two sets with an equal number of clusters.  The green horizontal lines represent the median values for each set, while the green shaded regions indicate the range that encompasses 50\% of the clusters above and below the median.} %
  \label{mass-proxy}
\end{figure*}

We tested our simulated clusters scaling relations against the observed clusters scaling relations reported by \citet[][]{2008A&A...482..451Z, 2009A&A...498..361P, Mantz_2010}. The first three panels in Fig. \ref{scaling relations} show the results of $L_{\rm{x}}$ -- $M_{500}$, $L_{\rm{x}}$ -- $T_{\rm{x}}$ and $T_{\rm{x}}$ -- $M_{500}$ scaling relations for the clusters selections given in  \citet[][]{2009A&A...498..361P}  in a redshift range of 0.05 < z < 0.164 (in red), \citet[][]{2008A&A...482..451Z}  spanning the range 0.14 < z < 0.3 (in yellow) and \citet[][]{Mantz_2010} containing clusters with z < 0.3 (in green). Even if this redshift threshold is small enough to keep evolution effects to a small level \citep[][]{Reichert_2011}, we corrected the data for possible redshift effects.

We calculated the luminosities, temperatures, and masses for our clusters in the $R_{500}$ radius using \textsc{Subfind} \citep[][]{2001MNRAS.328..726S, Dolag_2009, Saro_2006}. For the temperature, in order to focus on the ICM gas we performed a cutout above $10^5$K. The simulated data points for all clusters in our simulation are shown in gray, while the corresponding data points for our set of 46 clusters are plotted in blue.  %
All three figures show a great agreement between the observed clusters and the simulated clusters in our box. 

Furthermore, we examined the Compton-y - $M_{500}$ relationship for our simulated clusters. In this instance, we selected observational data from the \textit{Planck} database for clusters with redshifts below 0.1. For the simulated values, we generated Compton-y maps for each of the clusters in our simulations employing \textsc{Smac} \citep[][]{2005MNRAS.363...29D}, and computed the total Compton-y emission within $R_{500}$. Once more, our simulations faithfully reproduce the expected scaling relation presented in the observational data. Thus, we can conclude that our improved subgrid model presented in Sect. \ref{The SLOW Simulation Set} can also successfully replicate the crucial ICM characteristics such as $L_{\rm{x}}$ -- $M_{500}$, $L_{\rm{x}}$ -- $T_{\rm{x}}$ and $T_{\rm{x}}$ -- $M_{500}$ and $Y_{500}$ -- $M_{500}$ scaling relations.

\section{Results}
\label{Results}

\subsection{Local cluster and their replicas in SLOW}
\label{Local Cluster and their replicas in SLOW}

Table \ref{table:2} presents a comparison of the positions of local clusters and their simulated counterparts. The median separation between the observed cluster position and that of its replica in our simulation is of 25 Mpc/h, with only three exceeding a distance of 45 Mpc/h.

During the cluster identification process based on observed positions, two primary sources of errors come into play. Firstly, when using distance measurements derived from redshift for the positions of our galaxy clusters, we encounter uncertainties associated with the peculiar velocity of the cluster. To quantify this uncertainty, we referred to the study by \citet[][Fig. 1]{Dolag_2013}, where the authors presented histograms of peculiar velocities of galaxy clusters in a cosmological box across different mass bins. We adopted the 1 sigma error values for each mass bin presented in Fig. 1 of the mentioned paper to illustrate the typical peculiar velocity uncertainties, which have a mild dependence on mass. These resulting uncertainty values are displayed in column 8 of Table \ref{table:2}.

On the other hand, in simulations founded on peculiar velocities, uncertainties coming from distance moduli propagate to radial velocities and subsequently
extend further to uncertainties on the reconstructed density and total velocity field. Consequently, uncertainties in the observed distances (which are of around a 20\% of the distance value) are directly coupled with displacements of the three dimensional positions within the evolved simulation. 

In addition, a positional error originates from the creation of the constraints for the initial conditions, where scales below the linear threshold are typically non-constrained. However, this generally adds an error of 3 -- 4 Mpc, which is significantly smaller than the two errors discussed above.

Therefore the relative distances found for our simulated replicas, as presented in table \ref{table:2}, are reasonably small when compared to the above-discussed uncertainties, showing that we have a high-fidelity reproduction of these observed clusters in our simulation. Note that for half of the cluster replicas, the relative distances found are already almost fully covered when only considering the uncertainty of the observed cluster position based on their expected, typical peculiar velocity. 

\subsection{Detection significance}
\label{Detection Significance Subsection}

Originally, structure identification in reconstructions of the local Universe primarily relied on tracing the large-scale linear field \citep[][]{hoffman1993reconstruction, 1995ApJ...449..446Z, 1995velocity, 1998ApJ...492..439B} 
More recently, reproductions based on simulations have shifted their focus toward individual, collapsed structures, leading to a manual one-to-one structure and galaxy cluster identification \citep[e.g.,][]{2003ApJ...596...19K,2004JETPL..79..583D,2016MNRAS.458..900C, Sorce_2018, 2019MNRAS.486.3951S, 2020MNRAS.496.5139S, 2022MNRAS.515.2970S}{}{}. Recently, \citet{2023MNRAS.523.5985P} extended this type of  identification by associating well-defined probabilities to the cross-matched galaxy cluster pairs. Our current objective extends beyond the previous approach. We aim to identify collapsed, nonlinear structures at smaller scales within the larger-scale cosmological hydrodynamical box and assess their significance and the quality of their reproduction given the variety of multiwavelength observational data available for the individual galaxy clusters.  %

Therefore, once we have found counterparts for our galaxy cluster selection, we can consider the probability of such a finding
compared to a random position in a very large simulation box. Here, we can take the relative distance from the observed position,
the inferred mass of the galaxy cluster, or directly any observational signal that is typically used as a proxy for the cluster mass (like X-ray luminosity, temperature, etc.) directly into account.

In mathematical terms, we want to perform a hypothesis test, where we ask what is the p-value of our finding under the assumption of no difference compared to a null hypothesis, which in this case is a random simulation. This can give us a measure of how likely it is that our finding is in fact a product of the constraints in our ICs. Such a procedure is naturally only applicable for rare (e.g. relatively high-mass systems), but can be extended to smaller accompanying systems, see Seidel et al. in prep for details. Particularly, we performed this test using the four main cluster properties that we considered throughout this paper, which are $M_{500}$, $L\rm{x}_{500}$, $T\rm{x}_{500}$ and the $Y\rm{sz}_{500}$ signal. This time X-ray luminosities were also calculated using \textsc{Smac} in the 0.1 -- 2.5 keV band.
We subsequently explain the procedure for $M_{500}$, knowing that the analogous procedure applies to the rest of the properties. 

The steps are the following:
\begin{enumerate}

\item We select bins for $M_{500}$, ranging from $10^{14}$ M$_{\odot}$ to $10^{15}$ M$_{\odot}$ in steps of 10, giving a total amount of ten bins. Each of these bins will serve as a mass threshold for our cluster search.

\item We place a virtual observer at random positions in our cosmological box (a total of 1 million positions per mass bin),
which is equivalent to positioning ourselves in a random box.

\item We choose each of these mass bins and use it as a threshold for the cluster property. Then we search in a radius around the random position of the first cluster which has a value for $M_{500}$ larger than the %
threshold and we save this distance as our minimum radial distance. We perform this step using all bins as thresholds. 

\item We compute
a cumulative probability function for each of the mass bins,
in order to calculate the one-tailored p--value for each cluster.

\item Now we study our selected clusters in the light of these cumulative probabilities. We take the relative positions of our clusters and perform a spline interpolation between the cumulative probabilities of all bins at a fixed radius. By doing this we are able to estimate the exact probability of each of our clusters in terms of the $M_{500}$ value. 

\item We calculate the significance of the match as 1 -- p, where p is the p--value associated with that cluster. The significance values for each cluster are listed in Table \ref{table:2}.

\end{enumerate}

Figures \ref{significance1} and \ref{significance2} depict the cumulative distribution functions representing various bins of our four central properties. In these figures, blue points signify the observed clusters, accompanied by their associated property uncertainties. %
Correspondingly, the simulated counterpart values are denoted in red. Apart from the mass uncertainty, we also need to consider the distance uncertainty originating from peculiar velocities, as well as distance moduli, as discussed in Sect. \ref{Local Cluster and their replicas in SLOW}. Given the uncertainty in how the distance moduli errors affect our simulations, we have limited our analysis to the peculiar velocity uncertainty.
For clarity, we have graphically represented this uncertainty as a red line only for five %
clusters at different relative distances: Coma, Centaurus, A3532, A2107, and A496. This uncertainty applies however to all represented clusters. 

Probability values below the $-1\sigma$ threshold indicate a significance level exceeding 80\%. Consequently, we classify all clusters falling below this threshold as highly significant matches. It is worth noting that the significance of these clusters can exhibit notable variations when considering distance errors. In the worst-case scenario, even with distance errors {increasing the relative distance},
the significance for highly significant clusters remains at higher than 50\%. In contrast, under the best-case scenario, where the relative distance gets reduced, their significance can increase dramatically, surpassing the 99\% mark. %

Some of the lines for the values representing the higher-mass bins show unusual patterns, such as uneven spacing between bins or instances where two bins overlap. This is attributed to the inherent limitations of small number statistics within our cosmological box.
As we move into higher property value 
ranges, the occurrence of clusters with such values diminishes. Nevertheless, it is important to note that these irregularities do not have any relevant impact %
on our significance estimate.

Lastly, we would like to emphasize the exceptionally high significance values observed across all properties for several pivotal local galaxy clusters, notably including Virgo, Coma, Centaurus, Perseus, Fornax, and Norma, among many others (see Table \ref{table:2}). These significance values, hovering around 1.0, strongly indicate that the identification of these clusters is highly likely
attributable to the constraints we have imposed, rather than being a random occurrence.

\subsection{Properties' comparison}

A notable advantage
of our constrained local Universe simulations lies in our ability to perform direct one-to-one comparisons once we have successfully matched our clusters with their simulated counterparts. Figure \ref{property-distance} illustrates the property ratios between observed and simulated structures, in relation to their relative distances. The data points are color-coded based on their significance, with higher significance levels represented by blue, while lower significance levels are denoted by yellow.

In Fig. \ref{property-distance}a, we illustrate the mass ratio of SZ effect-derived mass within $R_{500}$ divided by the \textsc{Subfind}-estimated $M_{500}$. The dashed line in the plot represents the median value of the data, while the red-shaded region surrounding the median corresponds to the dispersion observed in the Compton-y -- $M_{500}$ scaling relations \citep[][see Fig. 5]{2011A&A...536A..11P}. The median value is 0.87
indicating that the SZ effect-derived mass tends to be lower than the simulated mass. This finding aligns with estimates of the hydrostatic mass bias.

The hydrostatic mass bias is 
relevant
for gas-derived mass estimations such as X-ray or SZ effect-derived masses, as it accounts for the fractional difference between the true mass and the cluster mass inferred using a gas proxy, assuming hydrostatic equilibrium.  This bias is estimated from simulations to be between 10\% and 20\% \citep{2009ApJ...705.1129L,2013ApJ...777..151L,2016ApJ...827..112B,2023A&A...670A..33S}.
In simpler terms, ICM-based mass estimates
tend to underestimate the actual mass by approximately a factor of 0.8-0.9, which closely matches the shift observed in the median of our mass ratios being $(1-b)=0.87$.

Furthermore, a majority of the clusters closely align with the median, and approximately half of them fall within the shaded region, %
which corresponds to the dispersion seen in the Compton-y -- $M_{500}$ scaling relation.

 We also notice that Fig. \ref{significance2}a shows a greater dispersion among clusters matched at smaller relative distances, whereas the dispersion tends to be more confined to the proximity of the shaded region for clusters matched at larger relative distances. This observation highlights a potential bias in our matching process. 

 Our matching process relies in its first steps on positions and masses. In the search process, sometimes we find a very massive galaxy cluster very close to the target cluster's observed position. This replica candidate has a high significance in terms of its mass as the relative distance between the observed and simulated cluster is small, pointing to the fact that its existence is very probably due to our constraints. Usually, after the complete selection procedure, that cluster will be selected as the corresponding cluster replica.

However, if we fail to identify a suitable counterpart in close proximity to the observed cluster, we expand our search radius.
This broader search increases the pool of potential candidates for a match. Therefore, the probability of finding a candidate with a very similar mass to the observed one becomes higher at these extended distances. Thus, the mass scatter becomes lower for these clusters. These trends are only slightly reflected for temperatures, luminosities, and Compton-y (see Figs. \ref{significance2}b,c,d).

In light of these results, we would like to discuss the replication quality of the subset of extensively studied clusters in the literature listed in Table \ref{table:3}. 

Coma demonstrates excellent agreement in terms of mass. Its simulated mass of 9.61 $\times 10^{14}$ M$_{\odot}$ aligns well with the X-ray derived mass of 9.95$^{+2.10}_{-2.99}$ $\times 10^{14}$ M$_{\odot}$, and it is just a bit higher than the SZ-derived mass. The simulated temperature of 7.06 keV also falls within the observed X-ray temperature range of 8.07 $\pm$ 0.29 keV. However, its luminosity and Compton-y values are higher by a factor of 1.5 and 3, respectively.

The Perseus cluster's simulated mass of 5.17 $\times 10^{14}$ M$_{\odot}$ is also in excellent agreement with its X-ray estimated mass of 6.08$^{+1.55}_{-2.85}$ M$_{\odot}$. The simulated temperature of 4.97 keV closely matches the reported X-ray temperature of 6.42 $\pm$ 0.06 keV.

Virgo with a simulated $M_{500}$ mass of 6.57 $\times 10^{14}$ M$_{\odot}$ and a virial mass of 7.12 $\times 10^{14}$ M$_{\odot}$, has a higher mass than the SZ estimated mass of 4.76$\pm$ 0.55 $\times 10^{14}$ M$_{\odot}$. However, considering various studies over recent years estimating Virgo's virial mass in the range of 6.0 - 8.0 $\times 10^{14}$ M$_{\odot}$, like \citet{1960ApJ...131..585D}, \citet{2014ApJ...782....4K} at 8.0 $\pm$ 2.3 $\times 10^{14}$ M$_{\odot}$, and \citet{2020A&A...635A.135K} at 6.3 $\pm$ 0.9 $\times 10^{14}$ M$_{\odot}$, \citep[see table 3 in ][for a longer list]{2023arXiv231002326L}{}{}, we still view the Virgo cluster's mass in the simulation as accurately reproduced.

Hydra also agrees in its mass estimation, with 2.64 $\times 10^{14}$M$_{\odot}$ in the simulation and $\rm 1.29^{+  0.44}_{-  0.55  }$ $\times 10^{14}$M$_{\odot}$ given by the X-ray observation. The temperature value lies also very close, from 3.37 keV in the simulation to 3.15 $\pm$ 0.05 keV from observations. It is important to note that these clusters have different dynamical states and likely underwent different formation paths, as discussed in Sect. \ref{A General View}

Clusters' replicas for Centaurus, AWM7, Fornax, and Norma deviate more from the observed quantities than the previously mentioned cluster replica. One key factor contributing to these differences in replication quality is the cluster's position in the sky. If we examine the positions of  Centaurus, AWM7, Fornax, and Norma, we find that they are located in close proximity to the "Zone of Avoidance" (as seen in Fig. \ref{Full sky plot}), where the extinction is severe. Consequently, the quality and quantity of observational data available for our constraints can be 
more limited. Perseus and Hydra, on the other hand, are better reproduced, even though 
they are near the Zone of Avoidance demonstrating the power of using peculiar velocities, which by tracing the potential can put constraints far beyond their actual sampling points. 

Compared to other simulations of the local volume constrained with galaxy distributions 
such as SIBELIUS-DARK \citep[][]{2022MNRAS.512.5823M}, we observe that the significant improvement in the replication
of crucial clusters like Perseus and Virgo \citep[][]{Sorce_2018} can now be extended to various other local clusters.
For instance, Perseus, with an observed X-ray-derived $M_{500}$ of $6.08\times 10^{14}$ M$_{\odot}$, was previously simulated with a mass of $1.87\times 10^{15}$ M$_{\odot}$ in SIBELIUS-DARK, whereas our replication yielded a closer mass of $5.17\times 10^{14}$ M$_{\odot}$.
Similarly for Virgo, its $6.57\times 10^{14}$ M$_{\odot}$ mass in our simulation lies in the range of $ 6-8\times 10^{14}$ M$_{\odot}$ reported in the previous mentioned studies by \citet{1960ApJ...131..585D}, \citet{2014ApJ...782....4K} and \citet{2020A&A...635A.135K}, while SIBELIUS-DARK contained a replica with a lower mass of  $2.7\times 10^{14}$ M$_{\odot}$ in $R_{500}$ and $3.5\times 10^{14}$ M$_{\odot}$ in $R_{200}$. 
Hydra replica is now also better reproduced in terms of mass, as we find a cluster replica with a mass of $2.6\times 10^{14}$ M$_{\odot}$, lying very well within the observed X-ray derived mass range of $\rm 2.50^{+  0.62}_{-  1.02  } \times 10^{14}$ M$_{\odot}$, in contrast to the $3.5\times 10^{14}$ M$_{\odot}$ mass value reported in SIBELIUS-DARK 
\citep[see Table B1 in ][for comparison]{2022MNRAS.512.5823M}. Indeed, when having a closer look at the quantity ratios in Fig. \ref{property-distance}, we see that the general mass scatter is low, pointing to a generally good cluster reproduction.

In this context, the cases of Norma and Centaurus are particularly intriguing, as they stand out as outliers in the panels depicted in Fig. \ref{property-distance}. Specifically, in the case of Norma, the simulated mass is approximately six times lower than the observed mass. As mentioned earlier, the proximity of the Norma cluster to the zone of avoidance and its association with the "Great Attractor" result in poorly constrained information about its mass and nature \citep[][]{1988ApJ...326...19L, Woudt_2007}. In addition, Norma exhibits an elongated structure in both observations \citep[as seen in][]{Woudt_2007} and our simulation. This characteristic may indicate a highly dynamic, accreting structure, further complicating the estimation of its true mass.

Norma shares its region of the sky with Hydra, Centaurus, and the Shapley clusters. Hydra and Centaurus form a cluster pair in the southern sky, positioned closer to us than Norma. While these two clusters are less massive, they are believed to play a central
role in shaping the 'Great Attractor' structure \citep[][]{Raychaudhury1989}. The challenge lies in their lower mass and their sky location, making it difficult to obtain high-quality observational data for accurate replication.

However, it is worth noting that the mass of Hydra is well reproduced, lying within the error region estimated from X-ray observations. The Centaurus cluster replica lies in the high mass end of the observed masses for this cluster. Indeed, its mass is 1.18 times higher than the dynamically estimated mass. Both structures if observed combined, constitute
a good replica of the Centaurus - Hydra pair. When considering the observed X-ray mass for Hydra ($2.5\times 10^{14}$ M$_{\odot}$) and the SZ-effect and dynamical derived masses for Centaurus ($1.23\times 10^{14}$ M$_{\odot}$ and $5.28\times 10^{14}$ M$_{\odot}$), the total pair mass falls within the range of $3.73\times 10^{14}$ to $7.78\times 10^{14}$ M$_{\odot}$. In our simulation, the combined mass of both clusters amounts to $8.85\times 10^{14}$ M$_{\odot}$, which lies only slightly higher than the observed range. Additionally, Fig. \ref{Centaurus-Hydra maps} depicts a density map of the Centaurus-Hydra region, with overlaid positions of observed galaxies. Galaxies directly assigned to
the Hydra and Centaurus clusters are marked in red, while those in the surrounding region are in orange. This figure illustrates our ability to accurately reproduce the relative positions of Centaurus and Hydra, as well as the structure of the surrounding cosmic web. Even more, evolving the local Universe forward in time
as presented by Seidel et al. in prep. shows that the Centaurus-Hydra pair merge. This indicates that although the individual clusters are less well replicated, the overall collapsing structure associated with them is well reproduced.

In conclusion, our simulations provide valuable insights into the properties of numerous key local galaxy clusters. However, there remains room for improvement, especially in accurately replicating 
even more clusters as well as structures situated near the zone of avoidance, where observational data is limited and replication could possibly be improved by the selection of certain, random realizations.

Improving the precision of these simulations poses several challenges: acquiring more comprehensive data, including currently sparsely sampled regions, refining the galaxy formation physics treatment within the simulations as well as improving observational mass estimates for better evaluation of similarity between simulated and observed clusters. Such improvements will be essential for gaining a deeper understanding of these complex sky regions.

\subsection{Observational tracers' reliability}

As previously discussed in the preceding
sections, it is important to acknowledge the presence of observational errors and biases, which can at times be challenging to precisely quantify. One of our central observational parameters is mass estimations.
It is noteworthy that these different mass estimation approaches yield significant discrepancies in their mass assessments for various clusters.%

A potential source of bias in mass estimation appears to be the distance of the clusters from the observer. To explore this possibility, we conducted tests to assess the behavior
of each mass estimation method in relation to distance. Initially, we conducted this analysis with respect to the simulated masses. Figure \ref{mass-proxy} illustrates the relationship between observed mass estimates and simulated masses in relation to distance. We categorized the clusters into two sets, one with closer members and the other with more distant members, each with an equal number of clusters. The median of both sets is represented by a green dashed line, while the green shaded region delineates the range encompassing 50\% of the clusters above and below the median.

Masses derived from X-ray emissions (Panel \ref{mass-proxy}b) and the SZ effect (Panel \ref{mass-proxy}c) exhibit minimal scatter around the median for both the closer and more distant cluster sets. The scatter shows no significant increase or decrease with cluster distance in these cases. In contrast, the scenario is quite different for dynamical masses (Panel \ref{mass-proxy}a), where the scatter notably increases for clusters at greater distances. %
        
Furthermore, we conducted comparative analyses among the various mass estimation methods while considering their relationship with distance, as depicted in Figs. \ref{mass-proxy} d, e, and f. When comparing dynamical masses against X-ray and SZ-effect derived mass estimations, dynamical masses consistently display a high scatter, which becomes even more pronounced for clusters located farther away, mirroring the pattern observed in Fig. \ref{mass-proxy}a. Conversely, the scatter between X-ray and SZ-effect derived mass estimations is significantly low, as evident in panel  \ref{mass-proxy}f. Nevertheless, the scatter between different observed
mass estimates is as large as the scatter between simulated and observed cluster counterparts. Taken together, this points to the fact that the level at which the constrained simulation reproduces the mass of the local galaxy clusters is at the level of the uncertainties in the observational mass estimates.

\section{Summary and conclusions}
\label{Discussion and Conclusions}

In this paper, we have presented the first results on cluster identification in our constrained, hydrodynamical cosmological simulation of the local Universe (SLOW), which includes cooling, star formation, and the evolution of super-massive black holes. We compiled a dataset of over 221 observed galaxy clusters and groups within the local volume covered by our simulation from existing literature,
collecting X-ray luminosities, temperatures, and Compton-y signals as well as inferred masses of these systems, when available. From this sample, we cross-identified 46 halos in the simulation, which are candidates to represent the according galaxy clusters within the local Universe. We then computed the significance of these associations based on different observables and compared the global properties between the observed and simulated counterparts. 

\begin{enumerate}

\item The galaxy clusters within the simulation generally follow the observed scaling relations between mass, temperature, X-ray luminosity, and SZ signal. This has allowed us for the first time to evaluate the cross identification between simulations and observations not only on the total mass but also directly on the full band width of observational signals from the ICM. Thereby such direct comparisons avoid the various biases inherited within the different methods to determine the total mass of galaxy clusters from observational proxies.

\item We compared the positions of cross-identified galaxy clusters between observations and simulations combined with the match of their various global properties (e.g., mass, total X-ray luminosity, temperature, or SZ signal) and computed the probability of finding a similar match in a random position within a large volume simulation. From this, we computed a significance of the match against the null hypothesis obtained from a random simulation volume.

\item This sample of clusters encompasses the 13 most massive clusters with a $M_{500}$ exceeding $2\times10^{14}M_\odot$, as identified by the SZ signal from Planck observations within a radius of approximately 300 Mpc. This highlights the success of our constrained simulation in precisely replicating the local galaxy clusters.

\item Based on the assessment of the significance of a match, only a small fraction cannot be distinguished from a random selection. On the contrary, even 18 of the matched clusters exhibited significance values exceeding 0.8 across all studied quantities. Notably, Virgo, Coma, and Perseus achieved a significance very close to 1.0 for all quantities with the smallest value being 0.98. %

\item Compared to other constrained simulations, we report a significant improvement in the number of identified replicas of crucial galaxy clusters in the local Universe, where masses now agree with their observational counterparts. Even for very prominent clusters such as Perseus, Virgo, and Hydra, we noticed that compared to simulations where the constraints are based on galaxy distributions, such as SIBELIUS-DARK, the mass is recovered better by the SLOW simulation.

\item We also investigated the impact of the observational distance uncertainty on our comparison. For the matched sample of galaxy clusters, the inferred displacement between the position in our simulation and the cluster position based on the observed redshift often reflects the uncertainties induced by the unknown peculiar velocity of the clusters. In the majority of cases, this displacement is smaller than the typical error in the distance modulus of the individual galaxies used for constructing the constraints for the initial condition. This assessment further demonstrates that our constrained simulation of the local Universe reproduces a large number of galaxy clusters that can be directly compared with their observational counterpart. 

\item Comparing the X-ray luminosity and temperature between observations and the simulated counterparts, the obtained width of the relative scatter is 1.7 and 1.5, respectively, for  
68\% of the clusters.
For the SZ signal, this scatter is of a factor 2.1. 
Interestingly, when comparing the sample median of the $M_{500}$ obtained from the simulations with the observed one inferred from the SZ signal, we would deduce a bias of $(1-b)=0.87$, which is in the range predicted by other simulations.

\item With our sample of cross-identified clusters, we compared the masses of the simulated and observed clusters using mass estimates based on different observables (this being: the SZ-effect derived mass, dynamical masses from Tully's galaxy catalog, and X-ray signal derived masses). We find that the scatter between different observed mass estimates is as large as the scatter between simulated and observed cluster counterparts. We neither find any strong correlation of the scatter with the distance of the cluster, nor with the significance assessment, except a tendency of the observational-inferred dynamical mass to have a higher scatter at larger distances with respect to other measurements as well as with respect to the simulations.
Overall this indicates that the level in which the constrained simulation is reproducing the mass of the local galaxy clusters is at the level of the uncertainties of the observational mass estimates.

\end{enumerate}

Overall, the current SLOW simulation of the local Universe faithfully replicates numerous fundamental characteristics for a sizable number of galaxy clusters within our local neighborhood. This is a first step in establishing an exciting avenue for cluster and cosmological research, where such simulations enable us to investigate the formation and evolution of individual galaxy clusters. This will allow us to study the physical processes governing the formation and evolution of galaxy clusters beyond the limitations inherited when using averaged populations toward marking the transition to test the effect of individual formation histories and environments.

\begin{acknowledgements}

We want to thank I. Khabibullin for his help and insightful comments on the X-ray-derived data appearing in our data tables.  The authors also thank the comments and discussions with all the members of this project. They thank the Center for Advanced Studies (CAS) of LMU Munich for hosting the collaborators of the LOCALIZATION project for a week-long workshop. This work was supported by the grant agreements ANR-21-CE31-0019 / 490702358 from the French Agence Nationale de la Recherche / DFG for the LOCALIZATION project. NA acknowledges support from the European Union’s Horizon 2020 research and innovation program grant agreement ERC-2015-AdG 695561 (ByoPiC, https://byopic.eu). KD and MV acknowledge support by the Excellence Cluster ORIGINS which is funded by the Deutsche Forschungsgemeinschaft (DFG, German Research Foundation) under Germany’s Excellence  Strategy – EXC-2094 – 390783311 and funding for the COMPLEX project from the European Research Council (ERC) under the European Union’s Horizon 2020 research and innovation program grant agreement ERC-2019-AdG 882679. MV is supported by the Fondazione ICSC National Recovery and Resilience Plan (PNRR), Project ID CN-00000013 "Italian Research Center on High-Performance Computing, Big Data and Quantum Computing" funded by MUR - Next Generation EU. MV also acknowledges partial financial support from the INFN Indark Grant. The calculations for the hydro-dynamical simulations were carried out at the Leibniz Supercomputer Center (LRZ) under the project pn68na. We are especially grateful for the support by M. Petkova through the Computational Center for Particle and Astrophysics (C2PAP). 

\end{acknowledgements}


%

\bibliographystyle{aa}
\bibliography{example} %



\appendix
\section{Galaxy Cluster Data Tables}

\begin{table}[ht!] %
{\small
\begin{threeparttable}
\caption{Positions, distance uncertainties, relative distance, and significances for our selected set of galaxy clusters. The observed positions are sourced from the Tully galaxy catalog and are represented in supergalactic coordinates. In cases where Tully's positions were unavailable, we extracted coordinates from NED \tnote{\textdagger} and subsequently transformed them into supergalactic coordinates. These clusters are denoted by a cross symbol. The distance uncertainties were determined through the consideration of peculiar velocity uncertainties, as elaborated upon in Sect.\ref{Detection Significance Subsection} for further details. Relative distance indicates the disparity between the estimated observational position and the position of the corresponding cluster in our simulation. Significances are assigned values ranging from 0 to 1, where higher numbers signify greater significance (i.e., a higher probability of deviation from a random simulation).}
\label{table:2}
\centering
\renewcommand{\arraystretch}{1.46} 
\setlength{\tabcolsep}{5.85pt}
\begin{tabular}{|c|c|c|c|c|c|c|c|c|c|c|c|c|c|c|c|c|c|c|c|c|c|}
\hline
\multirow{3}{*}{Name} & \multicolumn{3}{c|}{Simulations} & \multicolumn{3}{c|}{Observations}& \multirow{2}{*}{Distance} & \multirow{2}{*}{Relative}& \multicolumn{4}{c|}{Significance}  \\
 \cline{2-7}
 \cline{10-13}
     & SGX & SGY  & SGZ & SGX & SGY  & SGZ & Uncertainty & Distance& Lx$_{500}$&Tx$_{500}$ & Ysz$_{500}$ & M$_{500}$ \\
\cline{2-7}
\cline{10-13}
& \multicolumn{6}{c|}{[Mpc / h]} & \multicolumn{1}{c|}{[Mpc / h]} & \multicolumn{1}{c|}{[Mpc / h]} & \multicolumn{4}{c|}{(0 -- 1)} \\
\hline
        A85      & 50.84  & -143.10 & 25.88  & 46.39 & -155.60 & 1.76 &     15  & 27.53 & 0.99 & 0.96 &0.98&0.98\\ 
        \hline
        A119     & 73.61  & -117.28 & -24.93 & 57.59 & -119.46 & -1.67 &    15  & 28.32 & 0.98 & 0.98 &0.98&0.97 \\ 
        \hline
        A347     & 60.29  & -2.44   & -19.23 & 52.69 & -16.38 & -5.28  &    18  & 21.14  & 1.00 & 0.67 & 0.93& 0.20 \\ 
        \hline
        A496     & -13.28 & -94.52  & -83.46 & 24.79 & -51.18 & -83.46 &    17  & 57.69 & 0.69 & 0.61 &0.41&0.50 \\ 
        \hline
        A539     & 54.47  & 8.98    & -82.08 & 43.81 & -18.63 & -73.38 &    17 & 30.85 & 0.95 & 0.80 &0.93&0.86 \\ 
        \hline
        A576     & 120.30 & 39.98   & -14.59 & 98.80 & 57.53 & -38.41 &     17 & 36.57  & 0.70 & 0.60 &0.74&0.69 \\ 
        \hline
        A644     & -26.48 & 70.51   & -204.14 & -9.74 & 83.20 & -209.86 &   15 & 21.77  & 0.99 & 0.96 &0.98&0.98\\ 
        \hline
        A754     & -34.31 & 78.48   & -144.27 & -32.44 & 88.40 & -143.56 &  14 & 10.12 & 0.98 & 0.97 & 0.99 &0.98 \\ 
        \hline
        A1185    & 24.59  & 112.10  & -27.24 & 16.31 & 99.93 & -25.16 &     18  & 14.86  & 0.98 & 0.88 & 0.96&0.94\\ 
        \hline
        A1367    & -2.89  & 71.35   & -18.73 & -2.94 & 68.62 & -12.68 &     17 & 6.64 & 0.99 & 0.98 & 0.99&0.98 \\ 
        \hline
        A1644    & -97.26 & 64.92   & -5.74 & -102.00 & 107.93 & -10.76 &   15 & 43.57 & 0.93 & 0.70 & 0.85&0.76 \\ 
        \hline
        A1736    & -137.66 & 65.91  & -22.87 & -117.47 & 83.34 & -0.73 &    17 & 34.66 & 0.88 & 0.67 & 0.85&0.77 \\ 
        \hline
        A1795  \tnote{\textdagger}  & 6.26 & 201.72 & 29.19 & -9.57 & 174.19 & 59.27 & 15 &  43.74 & 0.30&  0.15 &0.35& 0.25\\ 
        \hline
        A2029 \tnote{\textdagger}   & -92.61 & 142.37 & 141.38 &  -94.65 & 162.38 & 127.79 & 14 &  24.27 &  0.99 &  0.95 & 0.98 &0.98\\ 
        \hline
        A2063    & -42.47 & 66.58   & 97.18 & -39.98 & 76.20 & 66.87 &      17 & 31.90 & 0.70 & 0.69 & 0.80&0.75 \\ 
        \hline
        A2065  \tnote{\textdagger}  & 7.25   & 138.68  & 107.11 & -9.07  & 161.56  &  134.02 & 15 & 38.91 & 0.94  &  0.89 &  0.94 &  0.90\\ 
        \hline
        A2107    & -28.27 & 71.39   & 83.16 & -18.25 & 92.93 & 89.98 &      17 & 24.71  & 0.73& 0.51 &0.60& 0.0.57\\ 
        \hline
        A2197/99    & -12.99 & 48.75   & 97.11 & 18.31 & 57.56 & 72.51 &       17 & 40.77 & 0.94 & 0.35 &0.69&0.61 \\ 
        \hline
        A2256    & 115.91 & 96.33   & 79.47 & 137.38 & 86.90 & 78.14 &      14 & 23.49  & 0.82 & 0.62 & 0.80&0.73\\ 
        \hline
         A2319 & 81.29 & 2.38 & 118.30 & 72.21 & 21.34 & 136.87  &     12 &  28.05 &  0.99 & 0.93 &  0.98 &  0.93 \\
        \hline
        A2572a  & 67.40 & -88.63 & 16.82 &  66.25 & -80.64 &  54.83  & 17 &  38.86 & 0.96 & 0.97 &0.90 &0.96\\ 
        \hline
        A2593    & 54.51  & -107.62 & 32.14 & 66.54 & -93.66 & 54.04 &      17 & 28.62  & 0.96 & 0.91 & 0.92&0.93\\ 
        \hline
        A2634    & 80.38  & -52.34  & 32.13 & 64.39 & -54.96 & 37.40 &      18 & 17.04  & 0.67 & 0.60 & 0.94&0.65\\ 
        \hline
        A2665    & 111.79 & -123.00 & 44.66 & 75.78 & -140.15 & 49.86 &     17 & 40.22  & 0.90 & 0.51 & 0.34&0.77\\ 
        \hline
        A2734    & -20.57 & -186.09 & 21.73 & -13.99 & -188.50 & 5.14 &     17 & 18.01 & 0.96 & 0.80 & 0.80 &0.88 \\ 
        \hline
        A2877 \tnote{\textdagger}   & -8.07 & -76.12 & -0.78 &  -21.18 & -65.46 &  -16.88 &  18 &  23.34 &  0.54&  0.50 &  0.96 & 0.57 \\ 
        \hline
        A3158    & -49.26  & -108.90 & -119.25 & -79.02 & -135.57 & -117.83  &      15 & 39.99  & 0.95 & 0.60 &0.93&0.72\\ 
        \hline
        A3266    & -59.95  & -93.17  & -128.12 & -97.79 & -105.76 & -111.61  &      14 & 43.17 & 0.97 & 0.90 &  0.65 &0.96  \\ 
        \hline
        A3376    & -36.34  & -74.00  & -124.29 & -41.49 & -50.16 & -130.44 &        17 & 25.15 & 0.93 & 0.82 &0.78&0.88 \\ 
        \hline
        A3391 / 95    & -53.30  & -43.16  & -101.67 & -78.28 & -51.37 & -118.14 &   17 & 31.03 & 0.99 & 0.97 &  0.46 &0.97 \\ 
        \hline
        A3532    & -156.71 & 96.63   & -36.03 & -143.78 & 96.68 & -22.69 &          15 & 18.58 & 0.87 & 0.82 & 0.98 &0.82 \\ 
        \hline
        A3558 & -141.91 & 66.10 & -37.17 & -130.29 & 78.7 & -3.64 &           15 &  37.66 &  0.98 &  0.99 & 0.85 &  0.98 \\
        \hline
        A3571    & -142.64 & 54.10   & -33.88 & -109.39 & 58.57 & 4.39  &           15 & 50.90 & 0.89 & 0.45 &0.78&0.60\\
        \hline

\end{tabular}
    \begin{tablenotes}[para,flushleft] %
      \item[\textdagger] NED Database: https://ned.ipac.caltech.edu
      \newline
      \item[\textasteriskcentered] For close-by systems such as Virgo we expect the error from direct distance measurements to be substantially smaller. 
      \newline
    \end{tablenotes}
    \end{threeparttable}}
    \end{table}

\begin{table*} %
{\small
\begin{threeparttable}
\caption{Table \ref{table:2}. continued.}
\centering
\renewcommand{\arraystretch}{1.46} 
\setlength{\tabcolsep}{5.85pt}
\begin{tabular}{|c|c|c|c|c|c|c|c|c|c|c|c|c|c|c|c|c|c|c|c|c|c|}
\hline
\multirow{3}{*}{Name} & \multicolumn{3}{c|}{Simulations} & \multicolumn{3}{c|}{Observations}& \multirow{2}{*}{Distance} & \multirow{2}{*}{Relative}& \multicolumn{4}{c|}{Significance}  \\
 \cline{2-7}
 \cline{10-13}
     & SGX & SGY  & SGZ & SGX & SGY  & SGZ & Uncertainty & Distance& Lx$_{500}$&Tx$_{500}$ & Ysz$_{500}$ & M$_{500}$ \\
\cline{2-7}
\cline{10-13}
& \multicolumn{6}{c|}{[Mpc / h]} & \multicolumn{1}{c|}{[Mpc / h]} & \multicolumn{1}{c|}{[Mpc / h]} & \multicolumn{4}{c|}{(0 -- 1)} \\
\hline
A3581    & -84.87  & 62.76   & -23.92 & -58.22 & 36.91 & 10.01 &            17 & 50.29 & 0.09 & 0.10 &0.90&0.20 \\ 
\hline
A3667    & -156.83 & -111.30 & 66.72 & -133.15 & -101.45 & 47.00 &          14 & 32.36  & 0.71 & 0.92 &  0.39 &0.90\\ 
\hline
2A0335+096   & 74.52 & -60.85 & -58.17 & 70.15 & -55.59 & -62.44 &          18 & 8.07 & 0.99 & 0.99 & 0.98&1.00 \\ 
\hline
3C129 \tnote{\textdagger}& 55.55 & 11.33 & -28.56 & 56.58 & 2.80 & -23.92 & 18 & 9.76 &  0.97 &  0.92 &  0.99 & 0.93  \\ 
\hline
AWM7 \tnote{\textdagger}& 60.08 & -6.71 & -23.32 &  45.01 &  -11.89 &  -8.47 & 18 & 21.78 &0.26&0.34& 0.93& 0.38  \\ 
\hline
Centaurus/A3526 & -22.82 & 11.21 & -13.26 & -34.25 & 14.93 & -7.56 & 18 & 13.31  &1.00&1.00& 0.88&1.00  \\ 
\hline
Fornax/AS0373 & 4.47 & -6.59 & -21.52 & -1.69 & -10.74 & -8.69 & 19 & 14.83  &0.90 & 0.53& 0.98&  0.70 \\ 
\hline
Hydra/A1060\tnote{\textdagger} & -21.21 & 9.89 & -19.41 &   -25.63 & 21.99 &   -25.97  & 18 &  14.45  &  1.00 & 0.95 & 0.27&0.97\\ 
\hline
Norma/A3627 & -45.65 & -0.38 & -5.67 & -50.26 & -7.06 & 6.44 & 17 & 14.58  &0.94&  0.66 &0.99&0.73 \\ 
\hline
Perseus/A426 & 59.22 & 2.13 & -24.58 & 49.94 & -10.73 & -12.98 & 18 & 19.65   &0.99&0.98& 1.00&0.98 \\ 
\hline
Coma/A1656 & -2.93 & 82.27 & -9.14 & 0.48 & 72.79 & 10.59 & 14 & 22.15  &1.00&1.00&1.00&1.00 \\ 
\hline
Virgo & -3.60 & 10.36 & -1.64 & -3.48 & 14.86 & -2.21 & 16 * & 4.54 &1.00&1.00&1.00&1.00  \\ 
\hline

\end{tabular}
    \begin{tablenotes}[para,flushleft] %
      \item[\textdagger] NED Database: https://ned.ipac.caltech.edu
      \newline
      \item[\textasteriskcentered] For close-by systems such as Virgo we expect the error from direct distance measurements to be substantially smaller. 
      \newline
    \end{tablenotes}
    \end{threeparttable}}
    \end{table*}

\begin{table*}
{\small
\begin{threeparttable}
\caption{X-ray luminosities ($L_{500}$) and temperatures ($T_{500}$) were primarily obtained from \citet{Ikebe_2002}. If available we also used more recent data from \citet{Shang_2008}\tnote{\textdaggerdbl} and \citet{2011A&A...536A..11P}\tnote{\textasteriskcentered}. For clusters not included in these studies, the corresponding data were extracted from the BAX database\tnote{\textdagger}. The SZ-signal ($Y\rm{sz}_{500}$) and its derived mass ($M\rm{sz}_{500}$) were sourced from the Planck database PSZ1v2. X-ray masses ($M\rm{x}_{500}$) were extracted from \citet{Chen_2007} correcting for the differences in the value of H$_0$ assumed. Dynamical masses were retrieved from the Tully galaxy catalog and subsequently transformed to $M_{500}$ by correcting by the 12\% factor presented by \citet{Sorce_2016} and the conversion method outlined by \citet{2021MNRAS.500.5056R}. Simulated values for $L_{500}$, and $Y\rm{sz}_{500}$ were estimated using \textsc{Smac}, while the simulated $M_{500}$ and $T_{500}$ were estimated using \textsc{Subfind}.}
\label{table:3}
\centering
\renewcommand{\arraystretch}{1.46} 
\setlength{\tabcolsep}{5.5pt}
\begin{tabular}{|c|c|c|c|c|c|c|c|c|c|c|c|c|c|c|c|}
\hline
\multirow{4}{*}{Name} & Obs & Sim & Obs & Sim & Obs & Sim & \multicolumn{3}{c|}{Obs} & Sim  \\
\cline{2-11}
     & Lx$_{500}$ & Lx$_{500}$ & Tx$_{500}$ & Tx$_{500}$ & Ysz$_{500}$ & Ysz$_{500}$ & Mx$_{500}$ &  Msz$_{500}$ & Mdyn$_{500}$  & M$_{500}$ \\
\cline{2-11}
& \multicolumn{2}{c|}{[10$^{44}$ erg s$^{-1}$]} & \multicolumn{2}{c|}{[KeV]} &\multicolumn{2}{c|}{[10$^{-4}$ Mpc$^2$]} & \multicolumn{4}{c|}{[10$^{14}$ $\times$ M$_{\odot}$]}  \\

& \multicolumn{2}{c|}{(band: 0.1-2.4 keV)} & \multicolumn{2}{c|}{ (band: 0.1-2.4 keV)}  &\multicolumn{2}{c|}{} & \multicolumn{4}{c|}{}   \\

\hline
A85 & 5.08 $\pm$ 0.07  & 8.06 & $\rm 6.51^{+  0.16}_{-  0.23  }$ & 5.33 &0.68 $\pm$ 0.05 & 1.12 & $\rm 6.06^{+  1.18}_{-  2.63  }$  & 4.90 $\pm$  0.21 & -- & 6.64 \\
\hline
A119 & 1.66 $\pm$ 0.028 \tnote{\textasteriskcentered} & 4.81 &  $\rm 6.69^{+  0.24}_{-  0.28  }$  & 5.59 & 0.33 $\pm$ 0.04 & 0.92 &  $6.73_{-1.94}^{+0.90}$ &3.34 $\pm$  0.22 & 2.83 & 5.82  \\
\hline
A347 & --  & 7.86 & -- & 0.40 & -- & 0.004 & --  &--& 3.19 & 9.16  \\
\hline
A496 & 2.19 $\pm$ 0.026  & 4.07 & 4.59 $\pm$ 0.10  & 4.44 & 0.23 $\pm$ 0.03 & 0.54 & $3.61_{-1.58}^{+0.67}$ & 2.74 $\pm$  0.17& 3.36 & 4.27  \\
\hline
A539 & 0.64 $\pm$ 0.013  & 4.13 &  $\rm 3.04^{+  0.11}_{-  0.10  }$  & 3.98 & -- & 2.32 & $\rm 2.01^{+  0.24}_{-  0.64  }$  &--&-- & 3.63  \\
\hline
A576 & 1.09 $\pm$ 0.12  & 2.77 & 3.83 $\pm$ 0.16  & 3.72 & 0.14 $\pm$ 0.02 & 0.13 & $3.46_{-1.79}^{+2.44}$  & 2.09 $\pm$  0.19 & 17.15 & 3.03  \\
\hline
A644 & 5.08 $\pm$ 0.09  & 5.51 & 6.54 $\pm$ 0.27  & 4.66 & 0.64 $\pm$ 0.06 & 0.76 &$\rm 6.31^{+  1.61}_{-  2.86  }$& 4.70 $\pm$  0.23 &--& 5.10  \\
\hline
A754 & 2.36 $\pm$ 0.07  & 2.25 & 9.00 $\pm$ 0.35 & 2.68 & 1.17 $\pm$ 0.06 & 0.13 &$\rm 10.39^{+  3.30}_{-  4.82  }$ & 6.68 $\pm$  0.20 & 6.65 & 1.91  \\
\hline
A1185 & 0.37 \tnote{\textdagger} & 3.11 & $\rm 3.9 ^{+  2.00}_{-  1.10  }$ \tnote{\textdagger} & 2.53 & 0.06 $\pm$ 0.02 & 0.14  &--&1.27 $\pm$  0.19 & 5.77 & 2.03  \\
\hline
A1367 & 0.68 $\pm$ 0.009 & 1.11 & 3.55 $\pm$ 0.08  & 2.19 & 0.1 $\pm$ 0.02 & 0.06 &$\rm 5.56^{+  0.83}_{-  1.78  }$& 1.76 $\pm$  0.14 & 2.65 & 1.10\\
\hline
A1644 & 2.32 $\pm$ 0.2  & 5.02 &  $\rm 4.70 ^{+  0.90}_{-  0.70  }$ & 4.53 & 0.43 $\pm$ 0.04 & 0.66 &$\rm 5.50^{+  3.22}_{-  3.30  }$& 3.83 $\pm$  0.21 & 6.70 & 4.40 \\
\hline
A1736 & 1.90 $\pm$ 0.19  & 3.71 &  $\rm 3.68 ^{+  0.22}_{-  0.17  }$ & 3.83 & 0.27 $\pm$ 0.05 & 0.41 &$\rm 1.63^{+  0.46}_{-  0.52  }$& 2.95 $\pm$  0.27 & 8.89 & 3.32\\
\hline
A1795 & 4.02 $\pm$ 0.03  & 1.97 &  $\rm 6.17 ^{+  0.26}_{-  0.25  }$ & 2.94 & 0.6 $\pm$ 0.05 & 0.16 &$\rm 7.40^{+  2.89}_{-  4.11  }$&4.54 $\pm$  0.21 & -- & 2.06 \\
\hline
A2029 & 10.47 $\pm$ 0.11  & 6.32 & $\rm 7.93 ^{+  0.39}_{-  0.36  }$ & 4.65 & 1.27 $\pm$ 0.08 & 0.93 &$\rm 7.46^{+  2.47}_{-  3.87  }$&6.82 $\pm$  0.24 &--& 6.09  \\
\hline
A2063 & 1.31 $\pm$ 0.027  & 2.34 & 3.56 $\pm$ 0.16  & 3.61 & 0.13 $\pm$ 0.03 & 0.26 &$\rm 1.77^{+  0.18}_{-  0.44  }$&2.03 $\pm$  0.23 & 4.52 & 2.87 \\
\hline
A2065 & 3.49 $\pm$ 0.022 \tnote{\textasteriskcentered} & 4.68 & 5.37 $\pm$ 0.34 & 5.30 & 0.55 $\pm$ 0.06 & 0.89 &$\rm 8.39^{+  7.18}_{-  5.11  }$&4.30 $\pm$  0.26 &--& 5.78 \\
\hline
A2107 & 1.41 \tnote{\textdagger} & 1.72 & 4.00 $\pm$ 0.10 \tnote{\textdagger} & 2.26 & 0.12 $\pm$ 0.03 & 0.06 &--&1.86 $\pm$  0.22 & 6.85 & 1.33 \\
\hline
A2197/99 & 2.43 $\pm$ 0.07  & 5.01 & 4.28 $\pm$ 0.10  & 3.33 & 0.23 $\pm$ 0.02 & 0.30 &$\rm 3.22^{+  0.88}_{-  1.42  }$&2.78 $\pm$  0.13 & 6.89 & 3.10 \\
\hline
A2256 & 2.04 $\pm$ 0.13 \tnote{\textdaggerdbl} & 2.07 &  $\rm 6.83^{+  0.23}_{-  0.21  }$ \tnote{\textdaggerdbl} & 2.46 & 1.08 $\pm$ 0.05 & 0.12 &$\rm 9.09^{+  2.56}_{-  3.09  }$ &6.34 $\pm$  0.18 & 43.03 & 1.74 \\
\hline
A2319 & 15.78 \tnote{\textdagger} & 5.25 & $\rm 8.84^{+  0.18}_{-  0.14  }$ \tnote{\textdagger} & 4.78 & 1.90 $\pm$ 0.08 &  0.75 & -- & 8.59 $\pm$ 0.22 & 27.20 & 4.57 \\
\hline
A2572a & 1.02 \tnote{\textdagger} & 4.81 & 2.5 $\pm$ 0.13 \tnote{\textdagger} & 6.40 & 0.15 $\pm$ 0.04 & 1.67 &--&2.14 $\pm$  0.27 & 3.06 & 7.57 \\
\hline
A2593 & 0.33 $\pm$ 0.048 \tnote{\textdaggerdbl}  & 4.16 & 3.10 $\pm$ 1.50 \tnote{\textdaggerdbl} & 4.63 & 0.14 $\pm$ 0.03 & 0.66 &--&2.09 $\pm$  0.26 & 3.32 & 4.56 \\
\hline
A2634 & 0.57 $\pm$ 0.016 & 0.55 & 3.45 $\pm$ 0.16  & 1.44 & --   & 0.08 &$\rm 3.38^{+  0.50}_{-  0.75  }$ &--& 2.40 & 6.27 \\
\hline
A2665 & 0.499 $\pm$ 0.076  \tnote{\textdaggerdbl} & 4.38 & 3.98 $\pm$ 0.07 \tnote{\textdaggerdbl} & 3.74 & 0.14 $\pm$ 0.04 & 0.45  &--&2.01 $\pm$  0.28& 1.40 & 4.09 \\
\hline
A2734 & 1.45 $\pm$ 0.058  & 2.78 &  $\rm 5.07^{+  0.36}_{-  0.42  }$ & 2.51 & 0.19 $\pm$ 0.04 & 0.13  &$\rm 3.61^{+  0.64}_{-  1.17  }$&2.38 $\pm$  0.26& 26.56 & 1.89\\
\hline
A2877 & 0.23 $\pm$ 0.005  & 0.97 & $\rm 3.50^{+  2.20}_{-  1.10  }$ & 2.06 & 0.05 $\pm$ 0.01 & 0.05  &$\rm 5.16^{+  5.05}_{-  2.84  }$&1.12 $\pm$  0.14&--& 1.20 \\
\hline
A3158 & 3.36 $\pm$ 0.09  & 4.98 &  $\rm 5.41^{+  0.26}_{-  0.24  }$ & 3.96 & 0.52 $\pm$ 0.04 & 0.46 &$\rm 4.31^{+  0.67}_{-  1.24 }$& 4.20 $\pm$  0.18 & 34.70 & 3.68 \\
\hline
A3266 &  5.17 $\pm$ 0.07  & 7.04 &  $\rm 7.72^{+  0.35}_{-  0.28  }$& 5.90 & 1.19 $\pm$ 0.06 & 1.70 &$\rm 14.43^{+  3.57}_{-  5.68  }$&6.71 $\pm$  0.18& 21.42 & 8.72 \\
\hline
A3376 & 1.27 $\pm$ 0.03  & 3.34 & $\rm 4.43^{+  0.39}_{-  0.38  }$& 3.64 & 0.17 $\pm$ 0.03 & 0.31 &$\rm 5.08^{+  1.16}_{-  1.49  }$ &2.27 $\pm$  0.20& 1.45 & 3.03 \\
\hline
A3391/95 & 1.57 $\pm$ 0.049  & 6.29 &  $\rm 5.89^{+  0.45}_{-  0.33  }$ & 6.02 & 0.14 $\pm$ 0.02 & 1.32 & $\rm 4.53^{+  0.55}_{-  1.27  }$ &2.04 $\pm$  0.18 &--& 7.02 \\
\hline
A3532 & 1.31 $\pm$ 0.038 & 1.40 &  $\rm 4.41^{+  0.19}_{-  0.18  }$& 2.64 & 0.3 $\pm$ 0.05 & 0.73 &$\rm 4.97^{+  0.88}_{-  2.13  }$ &3.09 $\pm$  0.26 & 2.62 & 1.56 \\
\hline
A3558 & 6.43 \tnote{\textdagger} & 6.69 & 4.78 $\pm$ 0.13 \tnote{\textdagger}& 7.65 & 0.55 $\pm$ 0.06 & 2.63 & -- & 4.41 $\pm$ 0.27 & 37.10 & 9.47 \\
\hline
A3571 & 4.72 $\pm$ 0.07 & 5.07 & $\rm 6.80^{+  0.21}_{-  0.18  }$ & 4.21 & 0.6 $\pm$ 0.05 & 0.50 &$\rm 6.57^{+  1.27}_{-  2.57  }$&4.67 $\pm$  0.21 & 4.29 & 4.09 \\
\hline
\end{tabular}
\begin{tablenotes}[para,flushleft]
\end{tablenotes}
\end{threeparttable}}
\end{table*}

\begin{table*} %
{\small
\begin{threeparttable}
\caption{Table \ref{table:3}. continued.}
\centering
\renewcommand{\arraystretch}{1.46} 
\setlength{\tabcolsep}{5.1pt}
\begin{tabular}{|c|c|c|c|c|c|c|c|c|c|c|c|c|c|c|c|}
\hline
\multirow{4}{*}{Name} & Obs & Sim & Obs & Sim & Obs & Sim & \multicolumn{3}{c|}{Obs} & Sim  \\
\cline{2-11}
     & Lx$_{500}$ & Lx$_{500}$ & Tx$_{500}$ & Tx$_{500}$ & Ysz$_{500}$ & Ysz$_{500}$ & Mx$_{500}$ &  Msz$_{500}$ & Mdyn$_{500}$  & M$_{500}$ \\
\cline{2-11}
& \multicolumn{2}{c|}{[10$^{44}$ erg s$^{-1}$]} & \multicolumn{2}{c|}{[KeV]} &\multicolumn{2}{c|}{[10$^{-4}$ Mpc$^2$]} & \multicolumn{4}{c|}{[10$^{14}$ $\times$ M$_{\odot}$]}  \\

& \multicolumn{2}{c|}{(band: 0.1-2.4 keV)} & \multicolumn{2}{c|}{ (band: 0.1-2.4 keV)}  &\multicolumn{2}{c|}{} & \multicolumn{4}{c|}{}   \\
\hline
A3581 & 0.35 $\pm$ 0.017  & 1.63 &  $\rm 2.83^{+  0.04}_{-  0.02  }$ & 3.18 & 0.35 $\pm$ 0.017 & 0.90 &$\rm 0.70^{+  0.14}_{-  0.28  }$& (1.83 $\pm$ 0.04)  & 1.10 & 2.40 \\
\hline
A3667 & 5.66 $\pm$ 0.07  & 2.43 &  $\rm 6.28^{+  0.27}_{-  0.26  }$ & 4.98 & 0.9 $\pm$ 0.07  & 0.49 &$\rm 3.96^{+  0.39}_{-  0.86  }$ & 5.77 $\pm$  0.24 & 27.38 & 4.07 \\
\hline
2A0335+096 & 2.58 $\pm$ 0.02  & 2.81 &  $\rm 3.64^{+  0.09}_{-  0.08  }$ & 3.52 & -- & 1.34 &$\rm  2.09^{+  0.82}_{-  1.22  }$ &--&--& 2.78 \\
\hline
3C129 & 1.27 $\pm$ 0.12 & 1.61 &  $\rm 5.57^{+  0.16}_{-  0.15  }$ & 1.78 & -- & 0.04 &  $\rm 4.04^{+  1.69}_{-  1.75  }$ &--& --& 0.87\\
\hline
AWM7 & 1.20 $\pm$ 0.04 & 0.24 &  $\rm 3.70^{+  0.08}_{-  0.04  }$ & 1.43 & -- & 0.05 &$\rm 3.69^{+  0.91}_{-  1.69  }$ &--& --& 0.54\\
\hline
Centaurus/A3526 & 0.302 $\pm$ 0.025 & 5.07 &  $\rm 3.69^{+  0.05}_{-  0.04  }$ & 5.73 & 0.05 $\pm$ 0.01 & 1.09 &--&1.23 $\pm$  0.11& 5.28 & 6.24\\
\hline
Fornax/AS0373 & 0.04 $\pm$ 0.003  & 1.26 & $\rm 1.56^{+  0.05}_{-  0.07  }$& 0.85 & -- & 0.04 &$\rm 0.97^{+  0.33}_{-  0.41  }$ &--& --& 0.36 \\
\hline
Hydra/A1060 & 0.32 $\pm$ 0.017  & 4.83 & 3.15 $\pm$ 0.05 & 3.37 & -- & 1.45 & $\rm 1.87^{+  0.46}_{-  0.76  }$ &--& --& 2.61\\
\hline
Norma/A3627 & 2.04 $\pm$ 0.1  & 1.73 &  $\rm 5.62^{+  0.12}_{-  0.11  }$ & 1.23 & 0.19 $\pm$ 0.04 & 0.01 &--&2.55 $\pm$  0.28 & 8.89 & 0.47\\
\hline
Perseus/A426 & 9.33 $\pm$ 0.11  & 3.91 & 6.42 $\pm$ 0.06 & 4.97 & -- & 4.34 &$\rm 4.56^{+  1.16}_{-  2.14  }$ &--& 7.96 & 5.17\\
\hline
Coma/A1656 & 4.63 $\pm$ 0.11 & 7.04 &  $\rm 8.07^{+  0.29}_{-  0.27  }$ & 7.06 & 0.73 $\pm$ 0.05 & 2.31 &$\rm 7.46^{+  1.57}_{-  2.24  }$ &5.29 $\pm$  0.20 & 4.77 & 9.61\\
\hline
Virgo & 0.254  $\pm$  0.005 \tnote{\textdaggerdbl}&  4.67 & $\rm 3.67^{+  0.53}_{-  0.52  }$  \tnote{\textdaggerdbl} & 5.70 & -- & 1.08 & 0.83 $\pm$ 0.01 \tnote{a}  & 4.76$\pm$ 0.55 \tnote{b}  & 3.43 & 6.57\\
\hline
\end{tabular}
\begin{tablenotes}[para,flushleft]
\item[a] Virgo X-ray derived mass was extracted from \citet{2017MNRAS.469.1476S}
\newline
\item[b] Virgo SZ-effect derived mass extracted from \citet{2016A&A...596A.101P}
\newline
\end{tablenotes}
\end{threeparttable}}
\end{table*}


\end{document}